\documentclass[12pt]{iopart}

\usepackage{graphicx}  
\newcommand{\bra}{\langle}
\newcommand{\ket}{\rangle}

\newcommand{\bfsigma}{\mbox{\boldmath $\sigma$}}
\newcommand{\bfmu}{\mbox{\boldmath $\mu$}}

\newcommand{\bfJ}{\mbox{\boldmath $J$}}

\newcommand{\dsp}{\displaystyle}
\def\hatm{\hat{m}}
\def\hatq{\hat{q}}

\def\vec0{\mbox{\boldmath $0$}}

\def\P{$P$}
\def\H{\mbox{$H$}}
\def\3M{\mbox{$3M$}}
\def\2M{\mbox{$2M$}}
\def\M{\mbox{$M$}}
\def\SG{\mbox{$SG$}}

\begin{document}

\title[Statistical mechanical study of partial annealing]
{
Statistical Mechanical Study on a Neural Network Model 
with Time Dependent Interactions
}

\author{T Uezu$^1$, K Abe$^1$, S Miyoshi$^2$ and M Okada$^{3,4}$}

\address{$^{1}$ Graduate School of Sciences and Humanities,
Nara Women's University, Nara  630-8506}
\address{$^{2}$ Department of Electrical and Electronic Engineering, 
Faculty of Engineering Science,
Kansai University, Osaka, 564-8680}
\address{$^{3}$ Devision of Transdisciplinary Sciences, Graduate School of 
Frontier Sciences,
The University of Tokyo, Chiba, 277-8561}
\address{$^{4}$ RIKEN Brain Science Institute, Saitama, 351-0198}
\ead{uezu@ki-rin.phys.nara-wu.ac.jp}
\begin{abstract}
We study a neural network model in which both neurons and synaptic
interactions evolve in time simultaneously.
The time evolution of synaptic interactions is described
by a Langevin equation  including a Hebbian learning term, 
and a bias term which is  the interactions of the Hopfield model.
We  assume that synaptic interactions change much slower
than neurons 
and study the stationary states of synaptic interactions by the replica
method. We find that the order of the phase transition changes
from the second to the first and that the existence regions
of the Hopfield attractor and mixed states increase as
the coefficient  of the learning term  increases.
We also  study the AT stability of solutions and 
find that the temperature region
in which the Hopfield attractor is stable
increases as the learning  coefficient
 increases.
Theoretical results are confirmed by the direct numerical
 integration of the Langevin equation.
Further, we study the characteristics of the resultant
synaptic interactions by partial annealing and 
find that the stability of the attractor which emerges after
partial annealing is enhanced and those of the
coexistent attractors are reduced.
\end{abstract}

\maketitle

\section{Introduction}
In this paper,  we consider the double dynamics of neurons
and synaptic interactions.  In particular, we study
the case that the time evolution of neurons is so rapid
that synaptic interactions are considered to be constant
 when the average of correlations of neurons are calculated.
We study the stationary states of the system
by the replica method.  In this study, the replica number
$n$ can be any value and the $n \to 0$ limit is not
taken  unlike the usual replica method.
The case in which the synaptic interactions do not
change we refer to as ``quenched'', whereas  the case in which 
they change together with neurons we call ``annealing''.
The present case in which synaptic interactions change much
slower than neurons is intermediate and is called ``partial annealing''.
Such systems have been studied previously, e.g., the case
that synaptic interactions learn the Hebbian rule\cite{CPS93,PCS93,PS94},
the case that the replica number is negative\cite{Dotsenkoetal94},
and the case that synaptic interactions are divided into a 
hierarchy of several groups with adiabatically separated and
monotonically increasing time-scale\cite{UC02}.
In this paper, we study the effect of the Hebbian learning
in  partial annealing introducing the coefficient of the Hebbian
learning.  If the coefficient is negative, we can study
the case of unlearning.  In this paper,  we mainly study
the case of learning, and study the states of neurons 
and the resultant synaptic interactions when the system
reaches to the stationary state\cite{Abe08,Hara08}.
Similar  analyses are under investigation for the Mexican-hat
 type interactions\cite{Hara08} and for
the Amit model\cite{Kimoto08}.\par
In the next section, we give the formulation of the model
and study the saddle point equations, the AT stability and the
phase transitions.
In section 3, the results of numerical simulations are presented
and are compared with theoretical results.
In section 4, we study the nature of interactions generated
by partial annealing. 
In section 5, a summary and discussion are given.
In appendix A, we give  the  analysis of the AT stability compactly.
\section{Formulation}
\subsection{Model}
We consider $N$ neurons.  Let $\sigma_i$ represent the state
of $i$th neuron which takes values $\pm1$, $\sigma_i=1$ is
the firing state and  $\sigma_i=-1$ is the rest state.
We assume $J_{ij}$ evolves in time according to the following Langevin
equation\cite{PCS93}:
\begin{eqnarray}
&&\hspace{-0.7cm}\tau\frac{d}{dt}J_{ij}= \frac{1}{N}
\varepsilon\bra\sigma_i\sigma_j\ket_{sp}
+\frac{1}{N} K_{ij} -\mu J_{ij}+\eta_{ij}(t)\sqrt{\frac{\tau}{N}}, \nonumber\\
& & \hspace{3cm} i<j=1, \cdots, N \label{eq:Langevin}
\end{eqnarray}
The first term 
$\bra\sigma_i\sigma_j\ket_{sp}$ is the expectation value 
of the correlation between $i$th and $j$th neurons and
represents the Hebbian learning.
We assume that this is calculated by the canonical distribution
with the Hamiltonian $H$ with the instantaneous values of 
$J_{ij}$ at time $t$.
\begin{eqnarray}
\bra\sigma_i\sigma_j\ket_{sp} &\equiv& \frac{1}{Z_\beta}{\rm Tr}_{\{\sigma_i\}}e^{-\beta H}\sigma_i\sigma_j,\\
H(\{ \sigma_i\}) &=& -\frac{1}{2}\sum_{i\ne j}J_{ij}\sigma_i\sigma_j,\\
Z_\beta &=& {\rm Tr}_{\{\sigma\}}e^{-\beta H}.
\end{eqnarray}
Here, Tr$_{\{ \sigma \}}$ denotes the summation of all
configurations of neurons $\{ \sigma_i\}$.
$Z_\beta$ is the partition function of neurons.
$\varepsilon$ is the learning coefficient.
The second term $K_{ij}$ is the basic synaptic interactions,
to which $J_{ij}$ converges if the learning term and the external 
noise do not exist.
In this paper,  as $K_{ij}$ we take the Hopfield model
with $p$ patterns,
\begin{eqnarray}
K_{ij}=\frac{K}{\sqrt{p}}\sum_{\nu=1} ^{p}\xi_i ^\nu \xi_j ^\nu. \label{eq:Kij}
\end{eqnarray}
The third term $-\mu J_{ij}$ represents the relaxation 
and is introduced so that $J_{ij}$ does not diverge.
The last term $\eta_{ij}$ is a white Gaussian random
variable with the mean 0 and the following covariance:
\begin{eqnarray}
\bra\eta_{ij}(t)\eta_{kl}(t^\prime)\ket=
2\tilde{T}\delta_{ik}\delta_{jl}\delta(t-t^\prime).\label{eq:white}
\end{eqnarray}
Here, $\tilde{T}$ represents the strength of the noise,
``noise temperature''.  The coefficients $\frac{1}{N}$ 
and $\frac{1}{\sqrt{N}}$ are scaling factors so that the
system has non-trivial limit as $N \to \infty$.
By defining $\cal H$ as
\begin{eqnarray}
{\cal H} =-\sum_{i<j}K_{ij}J_{ij}+\frac{\mu
 N}{2}\sum_{i<j}J_{ij}^2-\frac{\varepsilon}{\beta}\ln Z_\beta, 
\label{eq:Hamiltonian}
\end{eqnarray}
the Langevin equation is rewritten as follows:
\begin{eqnarray}
\tau\frac{dJ_{ij}}{dt}=-\frac{1}{N}\frac{\partial{\cal H}}{\partial J_{ij}}+\eta_{ij}(t)\sqrt{\frac{\tau}{N}}.\label{eq:Langevin2}
\end{eqnarray}
In the stationary state of eq.(\ref{eq:Langevin2}),
the probability density $P(\{J_{ij}\})$ of the  synaptic interactions
 $\{J_{ij}\}$ is given by
\begin{eqnarray}
P(\{J_{ij}\}) &\propto& e^{-\tilde{\beta}{\cal H}},\label{eq:ProbJij}\\
 \tilde{\beta} &=& \frac{1}{\tilde{T}}.
\end{eqnarray}
Thus, the partition function $\tilde{Z}_{\tilde{\beta}}$
of the total system is expressed by
\begin{eqnarray}
\tilde{Z}_{\tilde{\beta}} &=& \int d\bfJ e^{-\tilde{\beta}{\cal H}} \nonumber\\
&=& \int d\bfJ Z_\beta^n
 e^{-\frac{\tilde{\beta}}{2}N\mu\sum_{i<j}J_{ij}^2+\frac{\tilde{\beta}K}{\sqrt{p}}\sum_{i<j}J_{ij}\sum_{\nu=1}^p \xi_i^\nu \xi_j^\nu}. \label{eq:Ztilde}
\end{eqnarray}
Here, $\dsp d\bfJ=\prod_{i<j}dJ_{ij}$ and 
$n=\varepsilon\frac{\tilde{\beta}}{\beta}$.
Now, we calculate $Z_\beta^n$ by the replica method
 regarding $n$ as an integer.
Introducing $n$ replicas $\dsp \sigma_i^1 , \sigma_i^2 ,
\ldots, \sigma_i^n$, $\dsp Z_\beta^n$ is expressed as
\begin{eqnarray}
Z_\beta^n &=& \prod_{\alpha=1}^n{\rm Tr}_{\{\sigma_i^\alpha\}}e^{-\beta
 H(\{\sigma_i^\alpha\})} \nonumber\\
&=& {\rm
 Tr}_{\{\sigma_i^\alpha\}}
e^{\beta\sum_{i<j}J_{ij}\sum_\alpha\sigma_i^\alpha
\sigma_j^\alpha+\beta
\sum_{\nu,i}h_\nu\xi_i^\nu\sum_\alpha\sigma_i^\alpha}. \label{eq:Z^n}
\end{eqnarray}
We define order parameters 
$m_\nu^\alpha$ and $q^{\alpha\beta}$ as
\begin{eqnarray}
m_\nu^\alpha &=& \frac{1}{N}\sum_i\sigma_i^\alpha\xi_i^\nu,
 \label{eq:mnualpha}\\
q_{\alpha\beta} &=&
 \frac{1}{N}\sum_i\sigma_i^\alpha\sigma_i^\beta. \label{eq:qalphabeta}
\end{eqnarray}
Then, we obtain the following expressions:
\begin{eqnarray}
\tilde{Z}_{\tilde{\beta}}
&=& \sqrt{\frac{2 \pi}{\tilde{\beta}}N\mu}\int
 \frac{iN}{2\pi}d\hat{q}_{\alpha\beta}  \int \frac{N}{2\pi i}d\hat{m}_\nu
 ^\alpha e^{N G},  \label{eq:Z_1}\\
G &=& G_1+ G_2 + G_3, \\
G_1 &=&
 \frac{1}{4\mu}\tilde{\beta}K^2
+\frac{\beta^2}{2\mu\tilde{\beta}}
\sum_{\alpha<\beta}q_{\alpha\beta}^2
+\frac{\beta^2n}{4\mu\tilde{\beta}} +\frac{\beta K}{2\mu\sqrt{p}}
\sum_{\alpha,\nu}(m_\nu^\alpha)^2+\beta\sum_{\alpha,\nu}h_\nu
 m_\nu^\alpha,\label{eq:G_1}\\
G_2 &=&
-\sum_{\alpha<\beta}\hat{q}_{\alpha\beta}q_{\alpha\beta}
+\sum_{\alpha,\nu}\hat{m}_\nu^\alpha m_\nu^\alpha,\label{eq:G_2}\\
G_3
&=&   \frac{1}{N}\sum_i \ln \{ \mbox{Tr}_{\{\sigma_i^{\alpha}\}}
e^{ \sum_{\alpha<\beta}\hatq _{\alpha \beta}
 \sigma_i ^\alpha \sigma_i ^\beta
- \sum_{\alpha, \nu}
\hatm _{\nu}^{\alpha}  \sigma_i ^\alpha \xi_i ^ \nu}\}\\
&=& \Big[\ln\big\{{\rm Tr}_{\{\sigma^\alpha\}}
e^{\sum_{\alpha<\beta}\hat{q}_{\alpha\beta}\sigma^\alpha\sigma^\beta
-\sum_{\alpha,\nu}\hat{m}_\nu^\alpha\sigma^\alpha\xi^\nu}\big\}\Big].
\label{eq:G_3}
\end{eqnarray}
Since we consider the finite number of patterns $p$ and
take $N \to \infty$ limit in this paper,
$\dsp N \gg 2^p$ holds. Therefore,
in the first equation of the expression $G_3$,
$\dsp \frac{1}{N}\sum_i$ can be replaced by the 
average over $\{\xi\}$.  We denote this average by $\dsp [\cdots]$
and get the second equality.\\
Now, we assume the replica symmetry (RS) as
\begin{eqnarray}
q_{\alpha\beta} &=& q,\nonumber\\
m_\nu^\alpha &=& m_\nu,\nonumber\\
\hat{q}_{\alpha\beta} &=& \hat{q},\nonumber\\
\hat{m}_\nu^\alpha &=& \hat{m}_\nu. \label{eq:replica symmetry}
\end{eqnarray}
Let $G_{i, RS}$ be  $G_{i}$ evaluated at the RS solution.
Then, we obtain the following expressions for 
$G_{1,RS}, G_{2,RS}$ and $G_{3,RS}$:
\begin{eqnarray}
G_{1,RS} & =&
 \frac{1}{4\mu}\tilde{\beta}K^2
+\frac{\beta^2}{\mu\tilde{\beta}}\frac{n(n-1)}{4}q^2+\frac{\beta^2
 n}{4\mu\tilde{\beta}}  +\frac{\beta K}{2\mu\sqrt{p}}n\sum_\nu
 m_\nu^2 \nonumber \\
&&
+\beta n\sum_\nu h_\nu m_\nu, \label{eq:G_1replica}\\
G_{2,RS}  & =&
-\frac{n(n-1)}{2}\hat{q}{q}+n\sum_\nu\hat{m}_\nu m_\nu, \label{eq:G_2replica}\\
G_{3,RS}  & =& -\hat{q}\frac{1}{2}n+\bigg[\ln\int
Dx\Big\{\cosh\Big(\sqrt{\hat{q}}x
-\sum_\nu\hat{m}_\nu\xi^\nu\Big)\Big\}^n\bigg]
+n\ln2,\label{eq:G_3replica} \\
 & &
Dx=\frac{dx}{\sqrt{2\pi}}e^{-\frac{x^2}{2}}.\nonumber
\end{eqnarray}
\subsection{Saddle Point Equations (SPEs)}
Defining $ G_{RS}=G_{1,RS}+G_{2,RS}+G_{3,RS}$, 
we obtain the following saddle point equations. Here, we put $h_\nu=0$ 
and define $\dsp J=\frac{K}{\mu\sqrt{p}}$ and  $\dsp
\kappa=\frac{\beta^2}{\mu\tilde{\beta}}$.
\begin{eqnarray}
&& \frac{\partial G_{RS}}{\partial q}=0 : 
\hat{q} = \kappa q,\label{eq:antennew1}\\
&& \frac{\partial G_{RS}}{\partial \hat{q}}=0 : 
q = \bigg[\int Dx\cosh^n\Xi\tanh^2\Xi\Big\{\int
 Dx\cosh^n\Xi\Big\}^{-1}\bigg],
\label{eq:antennew2}\\
&& \frac{\partial G_{RS}}{\partial m_{\nu}}=0 : 
\hat{m_\nu} = -\beta Jm_\nu,\label{eq:antennew3}\\
&& \frac{\partial G_{RS}}{\partial \hat{m}_{\nu}}=0 : 
m_\nu = \bigg[\xi^\nu\int Dx\cosh^n\Xi\tanh\Xi\Big\{\int
 Dx\cosh^n\Xi\Big\}^{-1}\bigg],
\label{eq:antennew4}\\
& & \Xi=\sqrt{\kappa q}x+\beta\sum_\nu Jm_\nu\xi^\nu.
\end{eqnarray}
In this paper, we study the case of  $p=3$ and abbreviate the SPEs as
\begin{eqnarray}
q & \equiv & \varphi (q, m_1, m_2, m_3),
\label{eq:anten-1}\\
m_\nu & \equiv & \psi_{\nu} (q, m_1, m_2, m_3), \ (\nu=1,2,3).
\label{eq:anten-2}
\end{eqnarray}
We find the following solutions of the SPEs.
\begin{center}
\begin{itemize}
\item  $P$ \,\, :  $q=0$ , $m_\mu=0$, paramagnetic solution,
\item  $SG$ : $q>0$ , $m_\mu=0$, spin-glass solution,
\item  $H$ \,\, : $q>0$ , $m_1\ne 0$ , $m_2=m_3=0$, Hopfield attractor,
\item  $2M$ : $q>0$ , $m_1=m_2\ne 0$ , $m_3=0$, mixed state with 2 patterns,
\item  $3M$ : $q>0$ , $m_1=m_2=m_3\ne 0$, mixed state with 3 patterns.
\end{itemize}
\end{center}
In this paper, we  analyze the Hopfield attractor ($H$), 
the mixed state with 3 patterns ($3M$) and spin-glass state (\SG).
We do not consider the mixed state with 2 patterns (\2M)
because it is expected to be unstable. 
\subsection{AT stability}
We study the AT stability of the RS solution\cite{AT78}.
The condition of the stability is that
the free energy increases when order parameters deviate
from those at the RS solution. 
The free energy per neuron $\tilde{f}_{\tilde{\beta}}$
is given by
\begin{eqnarray}
\tilde{f}_{\tilde{\beta}}
=-\frac{1}{N}\frac{1}{\tilde{\beta}}\ln\tilde{Z}_{\tilde{\beta}}
=-\frac{G}{\tilde{\beta}}.
\end{eqnarray}
We define small deviations from the RS solution by 
\begin{eqnarray*}
m_\nu ^\alpha &=& m_\nu+\epsilon_\nu ^\alpha,\\
q_{\alpha\beta} &=& q+\eta^{\alpha\beta},
\end{eqnarray*}
and expand $G$ up to the second order of deviations.
Then, $G$ is expressed as
\begin{eqnarray}
G &=& G_{RS}+\frac{1}{2}\sum_{(\alpha\nu)(\beta\mu)}\!\!\!{\cal
 G}_{(\alpha\nu)(\beta\mu)}\epsilon_\nu^\alpha\epsilon_\mu^\beta
 +\frac{1}{2}\sum_{(\alpha\nu)(\beta\gamma)}\!\!\!{\cal
 G}_{(\alpha\nu)(\beta\gamma)}\epsilon_\nu^\alpha\eta^{\beta\gamma}
\nonumber \\
&& +\frac{1}{2}\sum_{(\alpha\beta)(\gamma\delta)}\!\!\!{\cal
 G}_{(\alpha\beta)(\gamma\delta)}\eta^{\alpha\beta}\eta^{\gamma\delta},
 \label{eq:zure}\\
G_{RS} & = & G\mid_{q_{\alpha\beta}=q,m_\nu ^\alpha = m_\nu}. \nonumber
\end{eqnarray}
${\cal G}$ is called the Hessian.
The stability condition of the RS solution is that
all eigenvalues of ${\cal G}$ are negative.
We calculated all eigenvalues  for \H, \3M and \SG.
There exist 7 different kinds of eigenvalues
for \H \ and \3M, and 5 different kinds of eigenvalues 
for \SG.
Details are given in Appendix.

\subsection{Phase Transition}
In this subsection, we study the phase transition.  The second order
phase transition temperatures are  determined by the
following relations:
\begin{eqnarray}
T_{P\rightarrow H}^{2nd} & = & T_{P\rightarrow3M}^{2nd}
=\frac{K}{\mu\sqrt{p}}=J,\\
 T_{P\rightarrow SG}^{2nd} & = &
\frac{1}{\sqrt{\mu\tilde{\beta}}}=\sqrt{\frac{\tilde{T}}{\mu}},\\
 T_{SG\rightarrow H}^{2nd}& = & J
\bigg\{\Big(\varepsilon\frac{T_{SG\to H}^{2nd}}{\tilde{T}}-1\Big)q+1\bigg\},\\
 T_{SG\to 3M}^{2nd} & = & T_{SG\rightarrow  H}^{2nd}.
\end{eqnarray}
Since the second order phase transition temperature from \P \  to \H,
 $T^{2nd}_{P \to H}$, 
and that from \P \ to \3M, 
 $T^{2nd}_{P \to 3M}$, are equal, we denote them by 
 $T^{2nd}_{P \to M}$.
Here \M \ implies both of \H \ and  \3M.
Similarly, since  $T^{2nd}_{SG \to H}$
and $T^{2nd}_{SG \to 3M}$ are equal, we denote them by 
 $T^{2nd}_{SG \to M}$.
Next, we study the first order phase transition.
In this case, a new phase appears
  suddenly irrespective of the old phase from which the
transition  takes place.
Thus, we consider the following three phase transitions
 and obtain the equations to determine the phase transition temperatures.
\begin{enumerate}
\def\labelenumi{(\theenumi)}
\item  Transition to \H 
\begin{eqnarray}
&  & q= \varphi(q,m_1,0,0),\\
&& m_1  =  \psi_1(q,m_1,0,0),\\
&& \Big(1-\frac{\partial\varphi}{\partial
 q}\Big)\Big(1-\frac{\partial\psi_1}{\partial
 m_1}\Big)-\frac{\partial\varphi}{\partial
 m_1}\frac{\partial\psi_1}{\partial q}=0.\label{eq:anteisei}
\end{eqnarray}
\item Transition to  \3M
\begin{eqnarray}
&& q=\varphi(q,m,m,m),\\
&& m=\psi_1(q,m,m,m),\\
&& \Big(1-\frac{\partial\varphi}{\partial
 q}\Big)\Big(1-\frac{\partial\psi_1}{\partial
 m}\Big)-\frac{\partial\varphi}{\partial
 m}\frac{\partial\psi_1}{\partial q}=0.
\end{eqnarray}
Derivatives in the above equations are calculated as
\begin{eqnarray*}
\frac{\partial\psi_1}{\partial m} &=& \frac{\partial}{\partial
 m}\psi_1(q,m,m,m)\\
&=& \frac{\partial\psi_1}{\partial m_1}+\frac{\partial\psi_1}{\partial
 m_2}+\frac{\partial\psi_1}{\partial m_3}=\frac{\partial\psi_1}{\partial
 m_1}+2\frac{\partial\psi_1}{\partial m_2},\\
\frac{\partial\varphi}{\partial m} &=& \frac{\partial\varphi}{\partial
 m_1}+\frac{\partial\varphi}{\partial
 m_2}+\frac{\partial\varphi}{\partial
 m_3}=3\frac{\partial\varphi}{\partial m_1}.
\end{eqnarray*}
\item  Transition to $SG$
\begin{eqnarray}
&& q=\varphi(q,0,0,0),\\
&& \frac{\partial\varphi}{\partial q}=1.
\end{eqnarray}
\end{enumerate}
We solve the above equations numerically and 
determine the phase transition temperatures.\par
Hereafter, we fix $K=1$ and $\mu=1$ and change 
$T, \tilde{T}$ and $\varepsilon$.
\setlength{\unitlength}{1.0mm}
\begin{figure}[pthb]
\begin{picture}(130,70)
\put(0,5){\includegraphics[width=8.cm,clip]{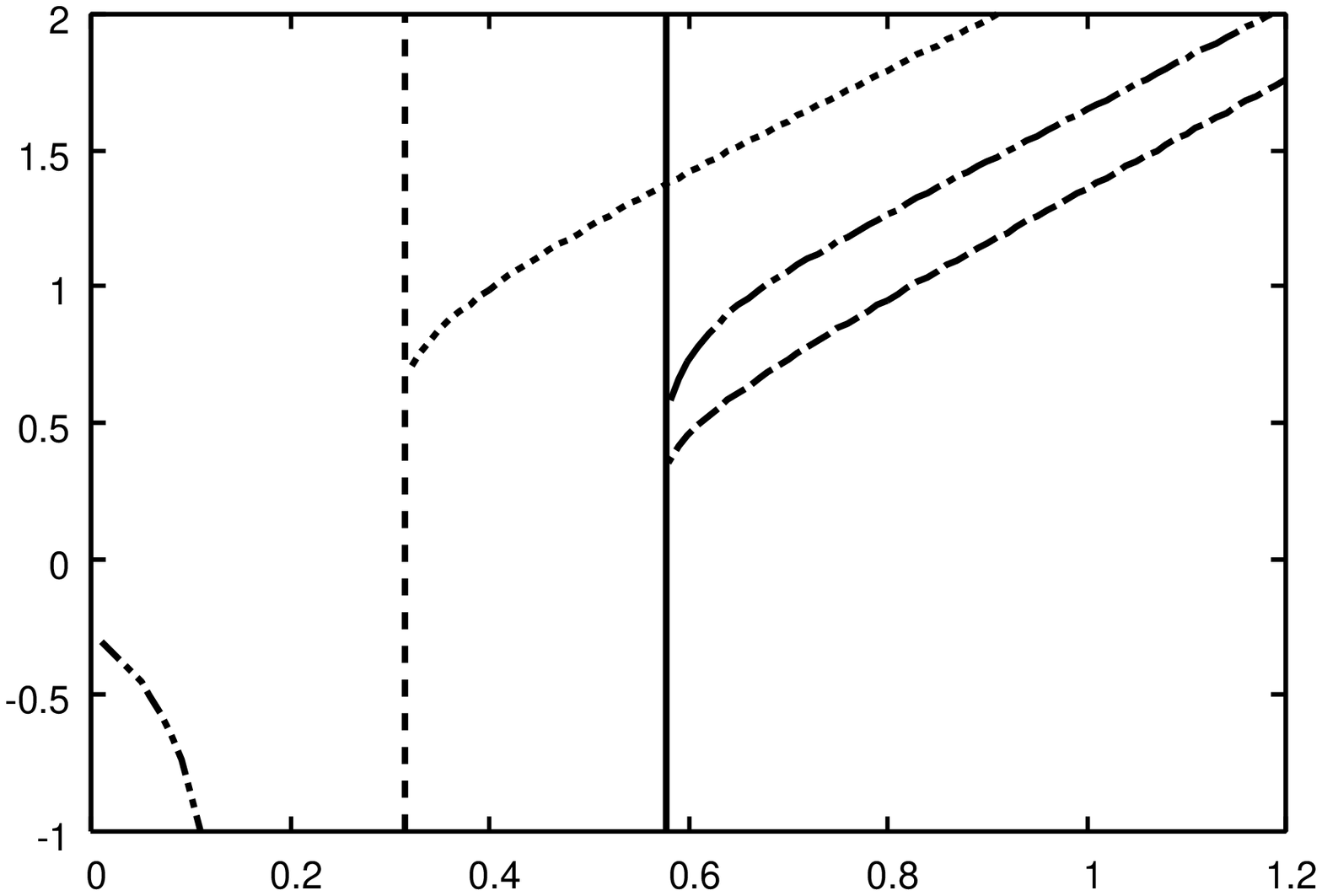}}
\put(38,0){ $T$}
\put(-3,34){$\varepsilon$}
\put(85,5){\includegraphics[width=8.cm,clip]{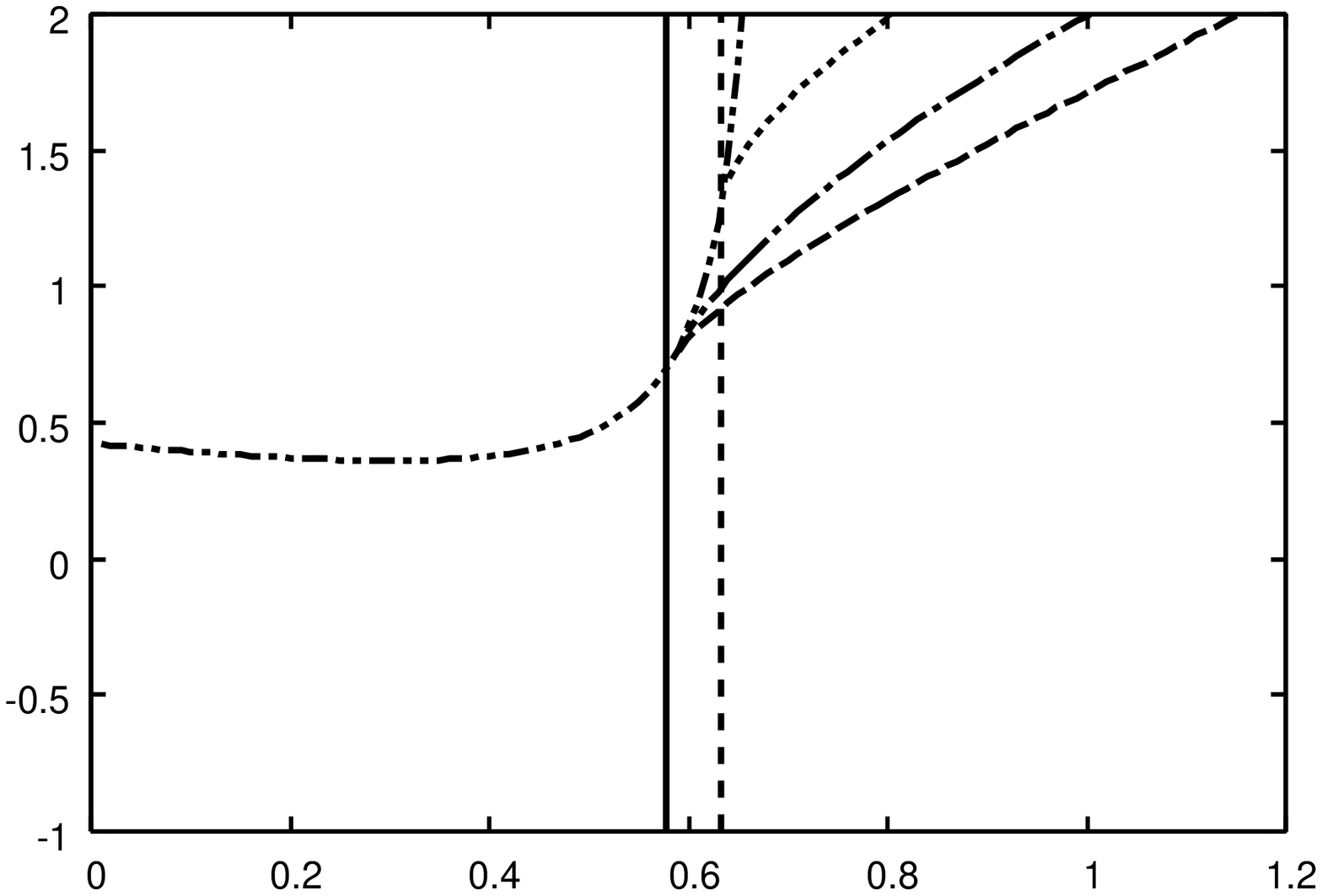}}
\put(124,0){ $T$}
\put(82,34){$\varepsilon$}
\end{picture}
\caption{Phase transition lines in
$T-\varepsilon$ plane.
Left panel: $\tilde{T}=0.1$. Right panel: $\tilde{T}=0.4$.
 $K=1.0$ and  $\mu=1.0$.
Solid curve: $T_{P\rightarrow M}^{2nd}$, 
short dashed curve:$T^{2nd}_{P \to SG}$,
dashed-dotted-dotted curve:$T^{2nd}_{SG \to M}$,
long dashed curve:$T^{1st}_{H}$,
dashed-dotted curve:$T^{1st}_{3M}$, 
dotted curve:$T^{1st}_{SG}$.
}
\label{fig.1}
\end{figure}
\setlength{\unitlength}{1.0mm}
\begin{figure}[hbt]
\begin{picture}(130,100)
\put(25,1){\includegraphics[width=12.cm,clip]{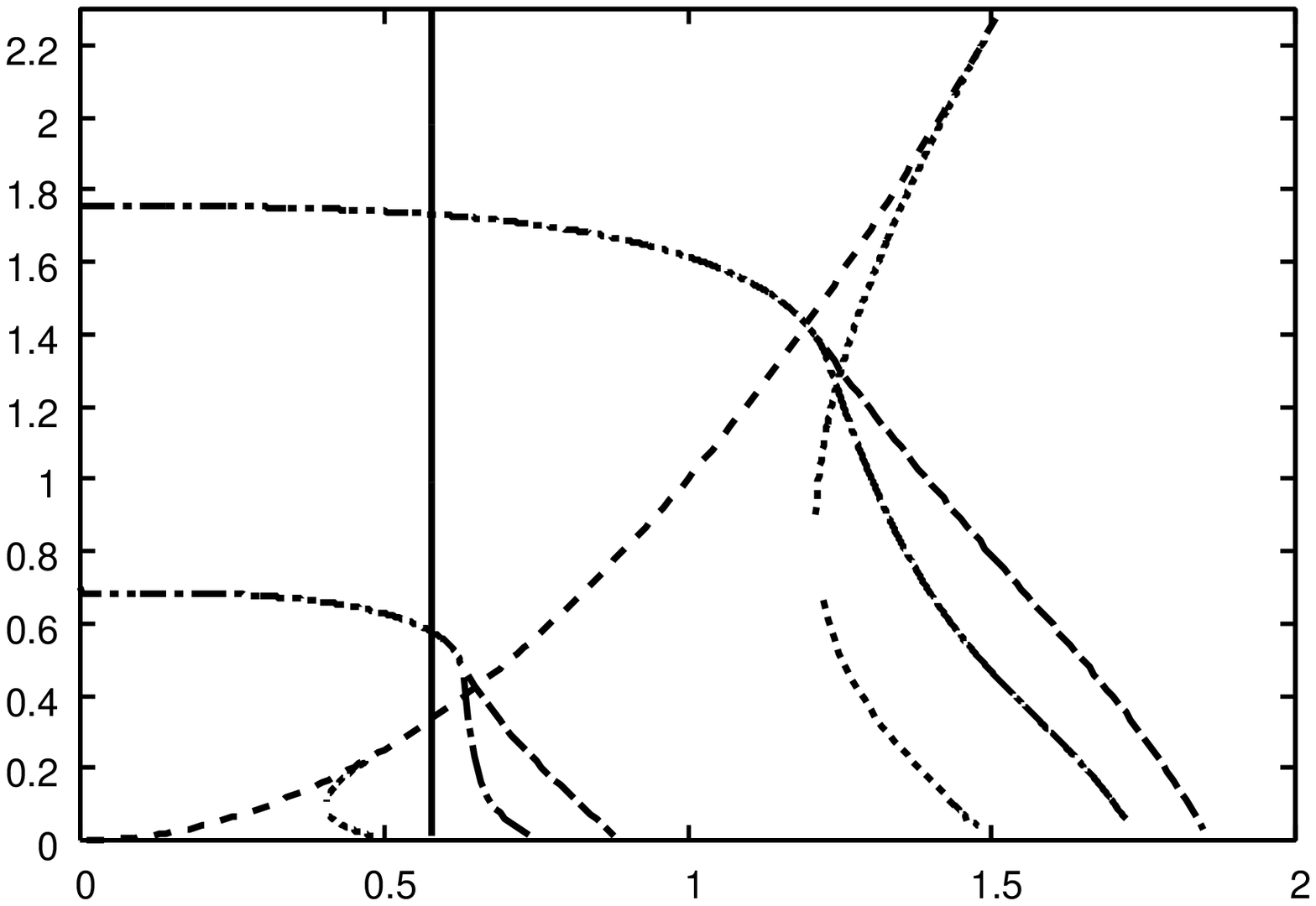}}
\put(87,-2){$T$}
\put(20,45){$\tilde{T}$}
\put(40,40){$\varepsilon=1.0$}
\put(120,40){$\varepsilon=3.0$}
\end{picture}
\caption{Phase transition lines in
$T-\tilde{T}$ plane for 
$\varepsilon=1.0$ and $\varepsilon=3.0$.
 $K=1.0$ and  $\mu=1.0$.
Solid curve: $T_{P\rightarrow M}^{2nd}$, 
short dashed curve:$T^{2nd}_{P \to SG}$,
dashed-dotted-dotted curve:$T^{2nd}_{SG \to M}$,
long dashed curve:$T^{1st}_{H}$,
dashed-dotted curve:$T^{1st}_{3M}$, 
dotted curve:$T^{1st}_{SG}$.
}
\label{fig.2}
\end{figure}
We display the phase transition lines 
in $T - \varepsilon$ plane for 
$\tilde{T}=0.1$ and 0.4 in Fig. \ref{fig.1},
and those in $T-\tilde{T}$ plane for 
 $\varepsilon=1.0$ and $\varepsilon=3.0$ in Fig. \ref{fig.2},
respectively.  
From these figures, we note the following.
There is a tendency that the order of the phase transition changes
from the second order to the first one as
$\varepsilon$ increases
for \H, \ \3M \ and \SG \ 
 or $\tilde{T}$ decreases for \H \ and \3M.
The interval of the temperature 
in which a solution exists  increases as 
$\varepsilon$ increases for \H, \ \3M \ and \SG \ 
or $\tilde{T}$ decreases for  \H \ and \3M.
In the right panel of Fig. \ref{fig.3}, we display the temperature
dependence of $q$ for 
$\tilde{T}=0.1$ and $ \varepsilon=1.0$.
For three cases of \H, \3M  and \SG, when the temperature is decreased,
a pair of solutions emerge at the temperature of the first order
phase transition, and two solutions are separated and consist of
the upper and lower branches as the temperature is further decreased.
Examining the AT stability, we find that the upper branch is 
stable and lower branch is unstable for \H \  and \3M, whereas
the both branches of \SG \ are unstable.
\setlength{\unitlength}{1.0mm}
\begin{figure}[pthb]
\begin{picture}(130,70)
\put(0,5){\includegraphics[width=9.cm,clip]{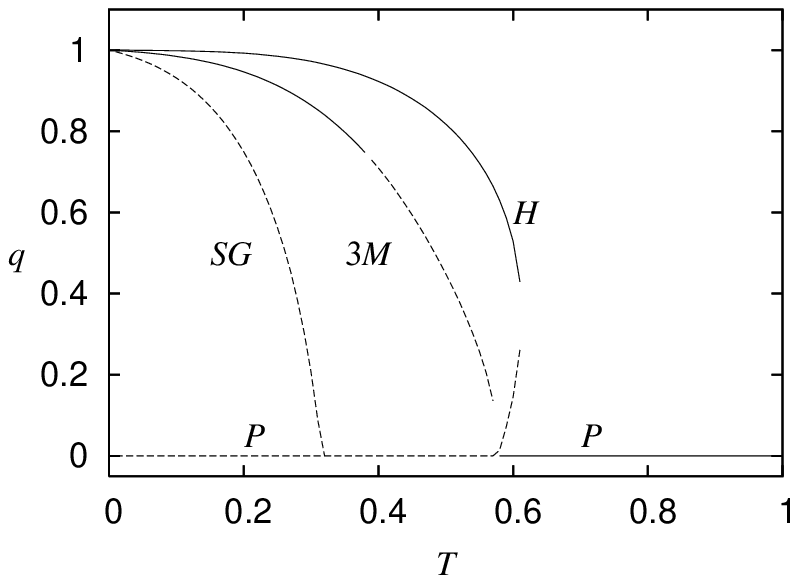}}
\put(85,5){\includegraphics[width=9.cm,clip]{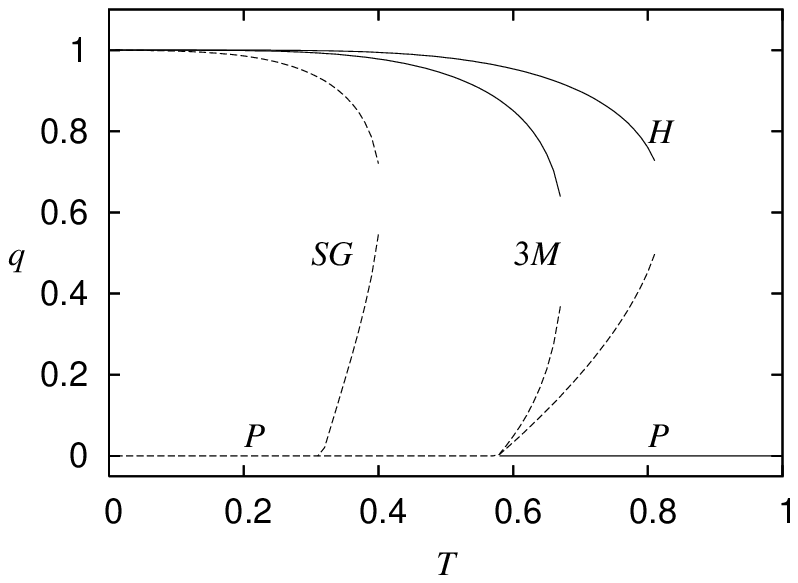}}
\end{picture}
\caption{
Temperature dependence of $q$. $\tilde{T}=0.1$.
Left panel: $\varepsilon=0.5$. Right panel $\varepsilon=1.0$.
 $K=1.0$ and  $\mu=1.0$.
Solid curve: stable solution. 
Dotted curve: unstable solution.
}
\label{fig.3}
\end{figure}
\setlength{\unitlength}{1.0mm}
\begin{figure}[pthb]
\begin{picture}(130,70)
\put(0,5){\includegraphics[width=9.cm,clip]{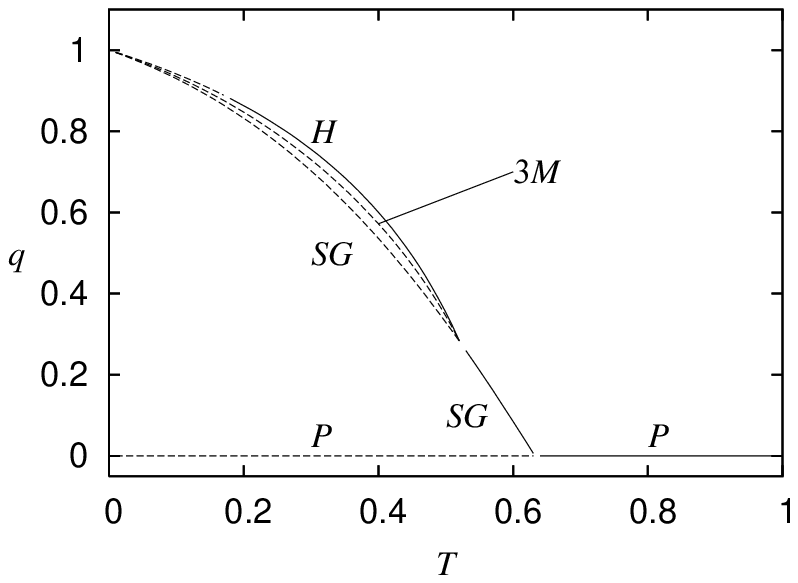}}
\put(85,5){\includegraphics[width=9.cm,clip]{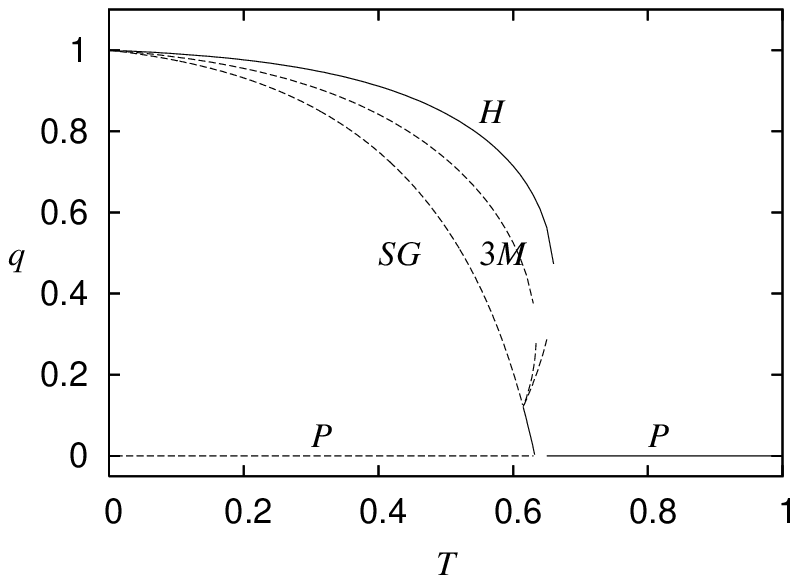}}
\end{picture}
\caption{
Temperature dependence of $q$. $\tilde{T}=0.4$.
Left panel: $\varepsilon=0.5$. Right panel $\varepsilon=1.0$.
 $K=1.0$ and  $\mu=1.0$.
Solid curve: stable solution.
Dotted curve: unstable solution.
}
\label{fig.4}
\end{figure}
\setlength{\unitlength}{1.0mm}
\begin{figure}[pthb]
\begin{picture}(130,200)
\put(25,98){\includegraphics[width=10.cm,clip]{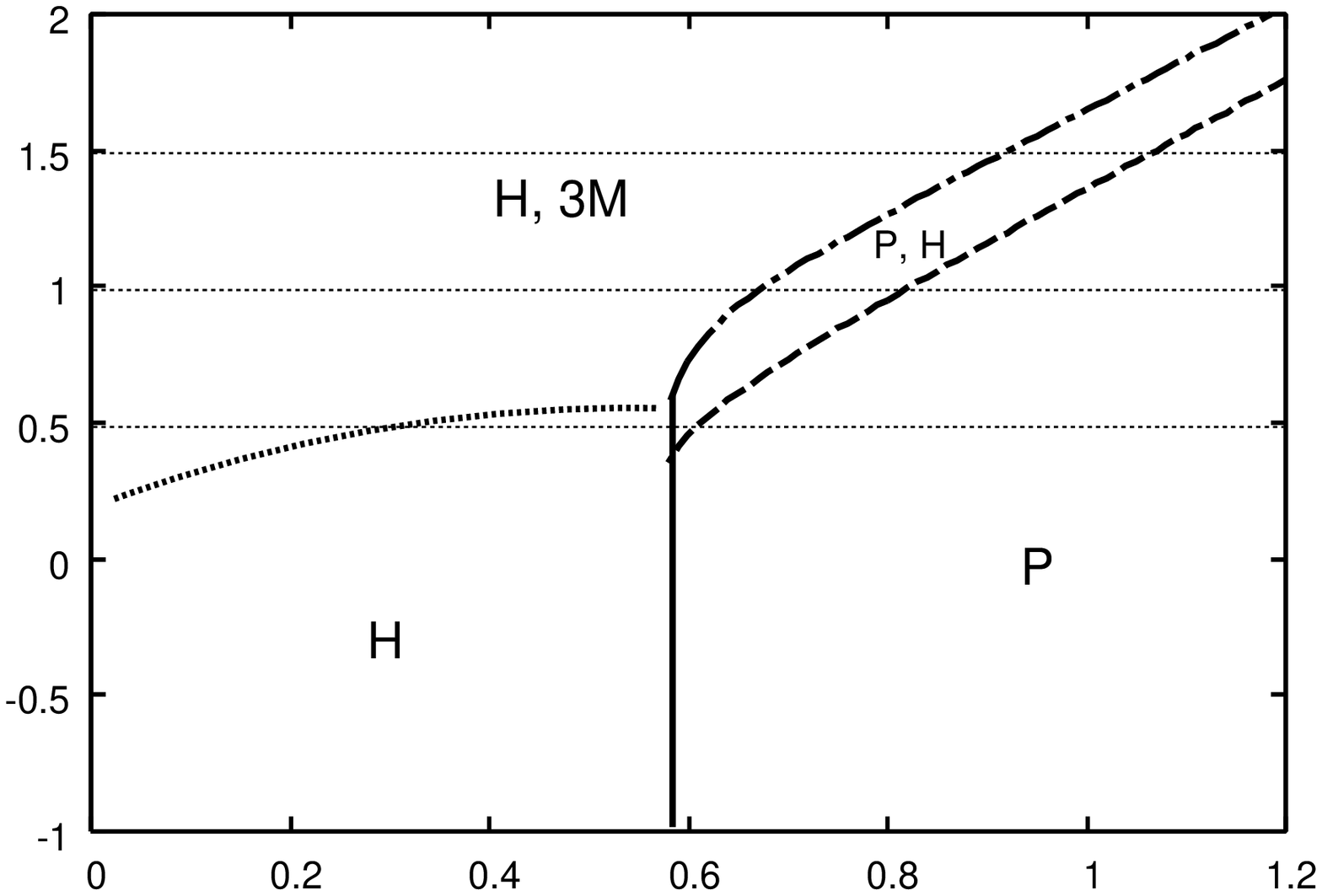}}
\put(75,95){$T$}
\put(20,135){$\varepsilon$}
\put(40,175){$\tilde{T}=0.1$}
\put(25,1){\includegraphics[width=10.cm,clip]{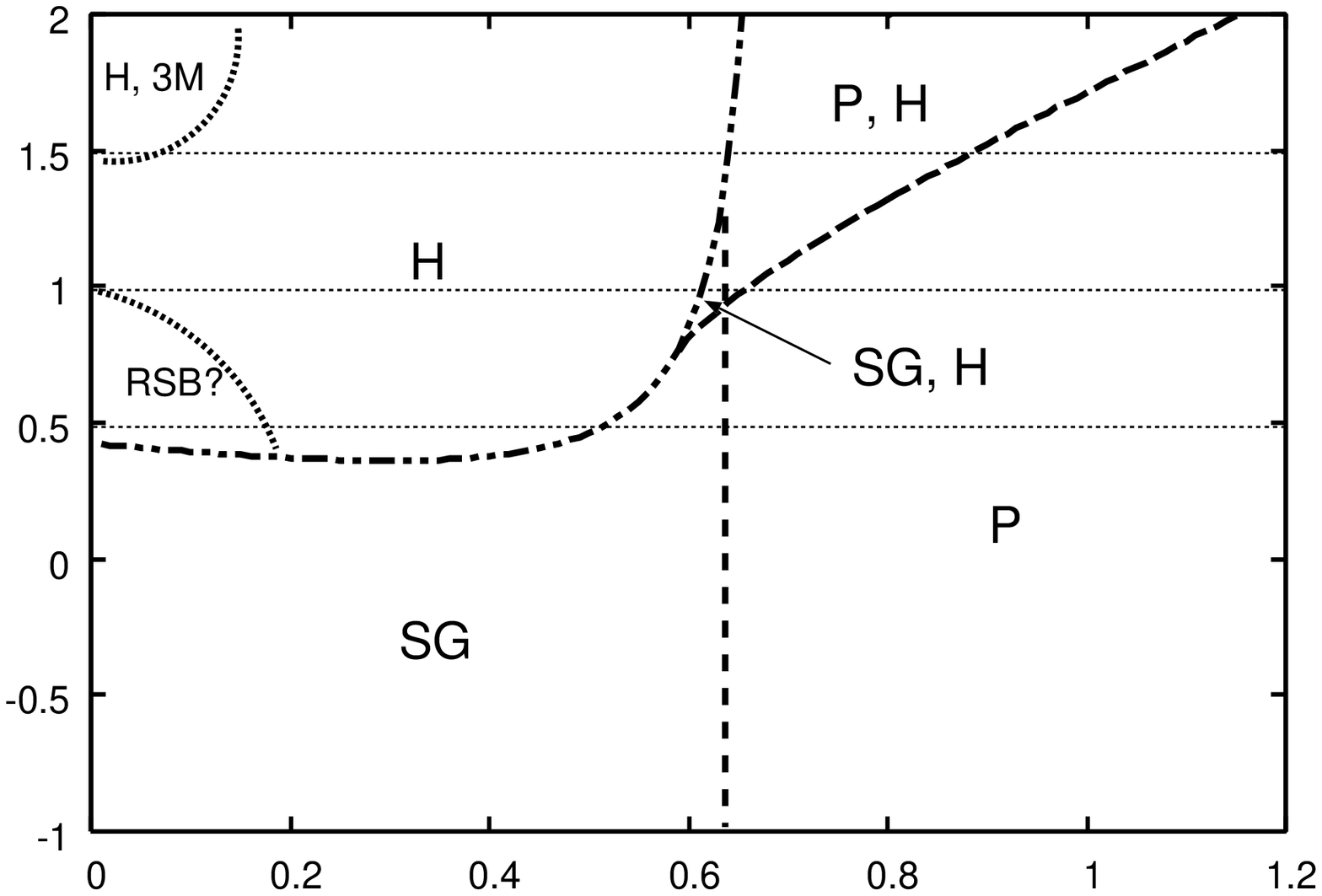}}
\put(75,-2){$T$}
\put(20,38){$\varepsilon$}
\put(40,78){$\tilde{T}=0.4$}
\end{picture}
\caption{
Phase diagram in $T-\varepsilon$ space. $K=1.0$, $\mu=1.0$, 
Upper panel: $\tilde{T}=0.1$,
Lower panel: $\tilde{T}=0.4$.
Solid curve: $T_{P\rightarrow M}^{2nd}$, 
short dashed curve:$T^{2nd}_{P \to SG}$,
dashed-dotted-dotted curve:$T^{2nd}_{SG \to M}$,
long dashed curve:$T^{1st}_{H}$,
dashed-dotted curve:$T^{1st}_{3M}$.
Dotted horizontal lines indicate the parameter where 
simulations were performed.  Dotted curves denote phase boundaries
which are not calculated theoretically but estimated by numerical
data available.
 }
\label{fig.5}
\end{figure}
As seen in the left panel of Fig. \ref{fig.3}
for  $\tilde{T}=0.1$ and $\varepsilon=0.5$,
\H \  emerges by the first order phase transition,
and \SG \ emerges by the second order one.
For \3M, it is difficult to determine the order of
the phase transition in this figure, but by Fig. \ref{fig.5}
we find that it is the second order.
The upper branch of \H \ is stable whenever it exists,
but \3M \ is stable only at low temperatures.\par
In Fig. \ref{fig.4}, we display the temperature dependence
of $q$ for $\tilde{T}=0.4$,
 the left panel is for $\varepsilon=0.5$
 and  the right  panel is for $\varepsilon=1.0$.
 We note that \H \ becomes unstable in low temperatures for
$\varepsilon=0.5$ and that there exist regions where \SG \  is stable.

The second order phase transition temperature 
$T_{P\rightarrow H}^{2nd}  =  T_{P\rightarrow3M}^{2nd}$
depends on $\tilde{T}$ but not on $\varepsilon$.
From this fact together with the fact that
when the first order phase transition takes place
 the upper branches
 of \H \ and \3M are stable for large $\varepsilon$ 
and small $\tilde{T}$ and the upper branch of \H \
is stable for large $\varepsilon$  and 
large $\tilde{T}$,
 we can understand
that the stable temperature region of the upper branch
increases when $\varepsilon$ is increased with fixed 
$\tilde{T}$.
In Fig. \ref{fig.5}, we display the phase diagram in $T-\varepsilon$
space taking into account the stability of the solutions.\par

\begin{table}
\caption{$\tilde{T}=0.1 , K=1 , \mu=1.0$.
Phase transition temperature of the
Hopfield attractor (\H)  and  the mixed
 state with 3 patterns (\3M), and 
the   ratio $r$ of the stable temperature
region for \3M \ to that for \H \ by
taking into account the AT stability for several values of
$\varepsilon$.
}
\begin{indented}
\item[]
\begin{tabular}{@{}llll}
\br
 & \H &  \3M  &  $\dsp r=\frac{\mbox{ stable region for \3M}}
{\mbox{stable region for \H}}$  \\
\mr
$\varepsilon=0$ & $0.58$  & $0.27$   & $0.47$  \\
$\varepsilon=0.5$ & $0.61$  & $0.38$    & $0.62$ \\
$\varepsilon=1.0$ & $0.83$ & $0.68$ & $0.82$ \\
$\varepsilon=1.5$ & $1.07$ & $0.92$ & $0.86$ \\
\br
\end{tabular}
\end{indented}
\label{soutenitemp}
\end{table}
In Table 1, we show the phase transition temperature of
\H, $T_c^{H}$, that of 
\3M, $T_c^{M}$,  and
 the ratio $r$ of the stable temperature
region for \3M to that for \H \ by
taking into account the AT stability for several values of
$\varepsilon$ and $\tilde{T}=0.1$.
We note that 
$T_c^{H}$, $T_c^{M}$ and  $r$ increases as 
$\varepsilon$ increases from 0.

\section{Simulations}
We performed direct integrations of the Langevin equation
(\ref{eq:Langevin}) fixing 
$\tau=1, K=1$ and $\mu=1$ and changing parameters 
$T, \tilde{T}$ and $\varepsilon$.
We used the Euler method with the time increment $\Delta t=0.1$.
We adopted the following procedure according to ref. 2).
When $N$ times of update by the Monte Carlo method are tried,
we call it 1 Monte Carlo step and denote it  by 1[MCS].
\begin{enumerate}
\item
Set an initial state of interactions $\bfJ(0)$.
\item  Update neurons $\bfsigma$ by $R_1$ [MCS].
\item  Calculate $\bra \sigma_i \sigma_j  \ket$ during the 
$R_2$ [MCS] update of neurons.
\item  Update $\bfJ$  by the Euler method.
\item Repeat 2 to 4 $R_3$ times.
\item Calculate averages of physical quantities during
the $R_4$ repetitions of 2-4.
\end{enumerate}
In this procedure, the total number of updates of
neurons is $(R_1+R_2)(R_3+R_4)$ [MCS].
As $R$s, we took $R_1=R_2=R_3=R_4=500$.
\setlength{\unitlength}{1.0mm}
\begin{figure}[thb]
\begin{picture}(130,110)
\put(7,0){\includegraphics[width=13.cm,clip]{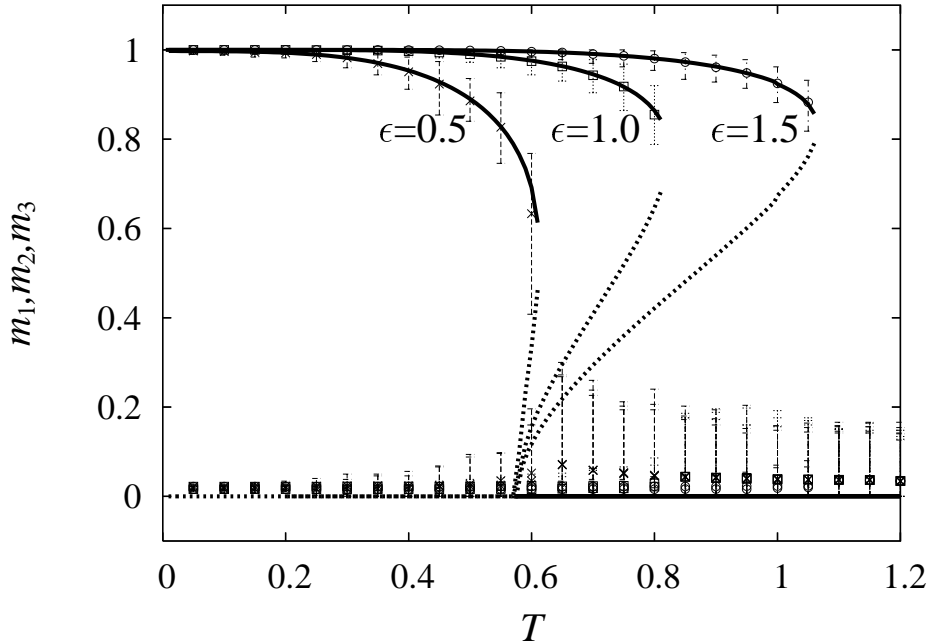}}
\end{picture}
\caption{Temperature dependence of $m_1 , m_2 , m_3$ 
for the Hopfield attractor
for $\varepsilon=0.5, \varepsilon=1.0$ and $\varepsilon=1.5$.
$\tilde{T}=0.1 , K=1.0$ and $\mu=1.0$.
Curves are theoretical results.
Solid curves are stable and dotted curves are unstable.
Symbols are simulation results for  $N=1000$.
}
\label{fig.6}
\end{figure}

\setlength{\unitlength}{1.0mm}
\begin{figure}[bht]
\begin{picture}(130,110)
\put(7,0){\includegraphics[width=13.cm,clip]{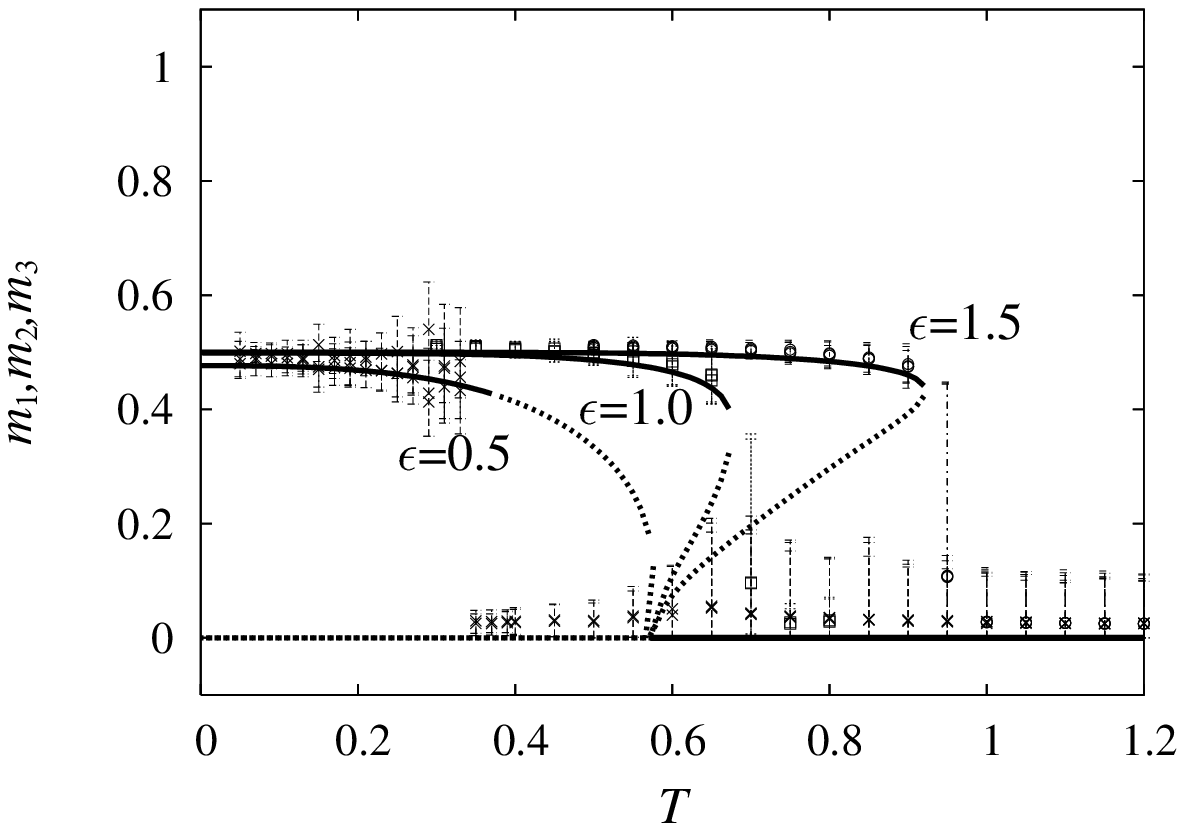}}
\end{picture}
\caption{Temperature dependence of $m_1 , m_2 , m_3$ 
for the mixed state 3M 
for $\varepsilon=0.5, \varepsilon=1.0$ and $\varepsilon=1.5$.
$\tilde{T}=0.1 , K=1.0$ and $\mu=1.0$.
 Curves are theoretical results.
Solid curves are stable and dotted curves are unstable.
 Symbols are simulation results for $N=2000$.
}\label{fig.7}

\end{figure}

First, 
we show the numerical results
together with theoretical ones for the Hopfield attractor 
in Fig.\ref{fig.6}.
Parameters are $\tilde{T}=0.1$, $\varepsilon= 0.5, 1.0, 1.5$
and $N=1000$.  Numerical results agree fairly well with the
upper branch solution obtained theoretically.
Next, the theoretical and numerical results are shown for the mixed state, 
in Fig.\ref{fig.7}.
Parameters are $\tilde{T}=0.1$, $\varepsilon=0.5, 1.0. 1.5$
and $N=2000$.
The agreement between simulations and the stable upper branch
 solution by theory are quite well, except for
two cases.
One is just above the critical temperature 
for $\varepsilon=1.0$ and  $\varepsilon=1.5$.
 In the paramagnetic phase, fluctuations are very large. 
The other is around the temperature where
the AT instability takes place for $\varepsilon=0.5$.
In this case, the phase transition takes place
at the temperature which is lower than the theoretical prediction.
Results in  both cases are  considered to be finite size effects.

\section{On the nature of interactions generated by partial annealing}
Here, we numerically study the interaction $\{ J_{ij} \}$
which appears when the system reaches to the stationary state.
There are two types of such interactions. One is 
$\{ J_{ij}^H \}$ for which \H \
emerges at the stationary state,  and the other is 
$\{ J_{ij}^M \}$ for which the \3M \ emerges.
Using these two interactions, we performed Monte Carlo simulations 
taking two initial conditions, \H \ and \3M.

\setlength{\unitlength}{1.0mm}
\begin{figure}[pthb]
\begin{picture}(130,70)
\put(0,5){\includegraphics[width=8.cm,clip]{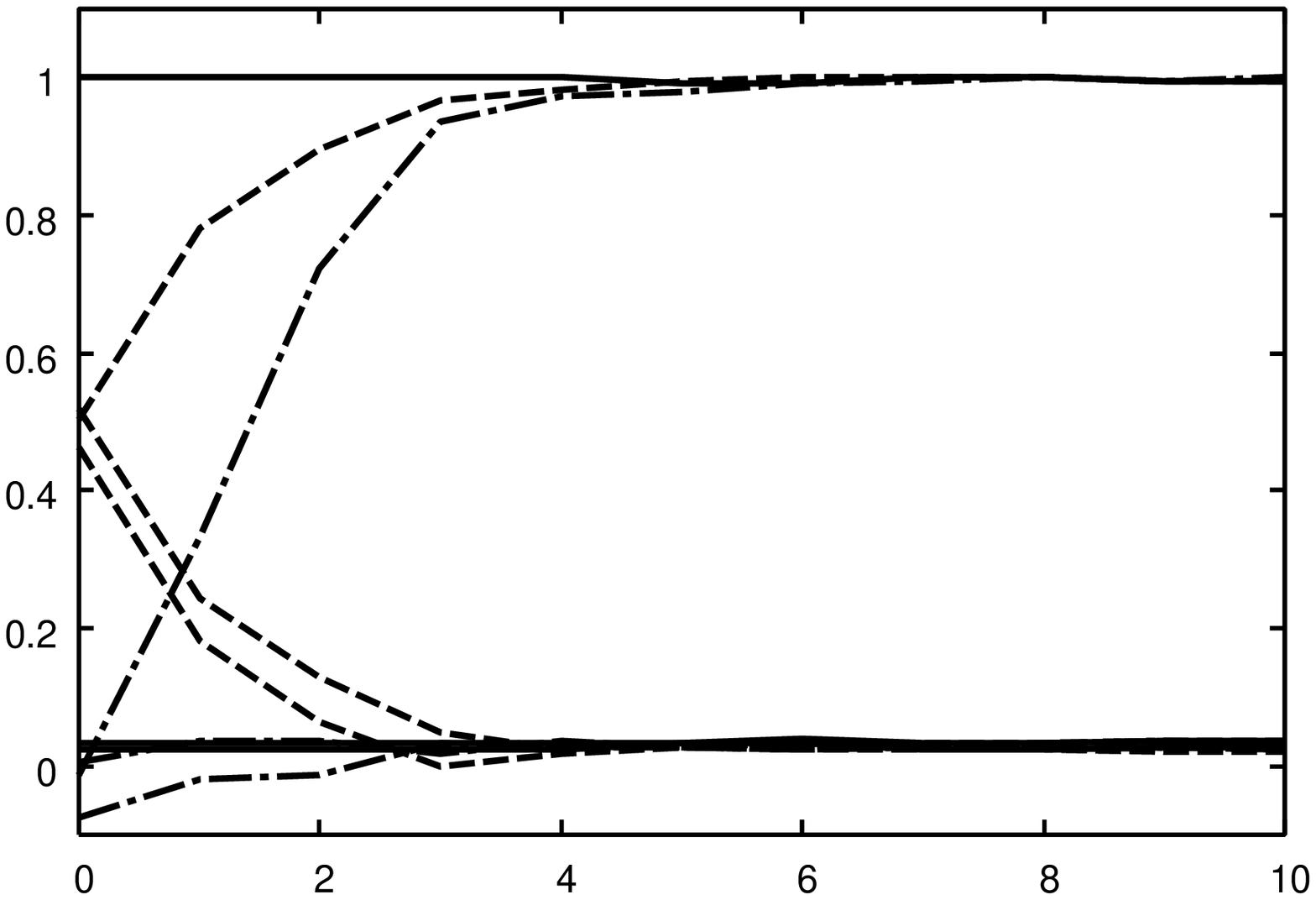}}
\put(35,0){ time[MCS]}
\put(-5,40){$m_3$}
\put(-5,34){$m_2$}
\put(-5,28){$m_1$}
\put(85,5){\includegraphics[width=8.cm,clip]{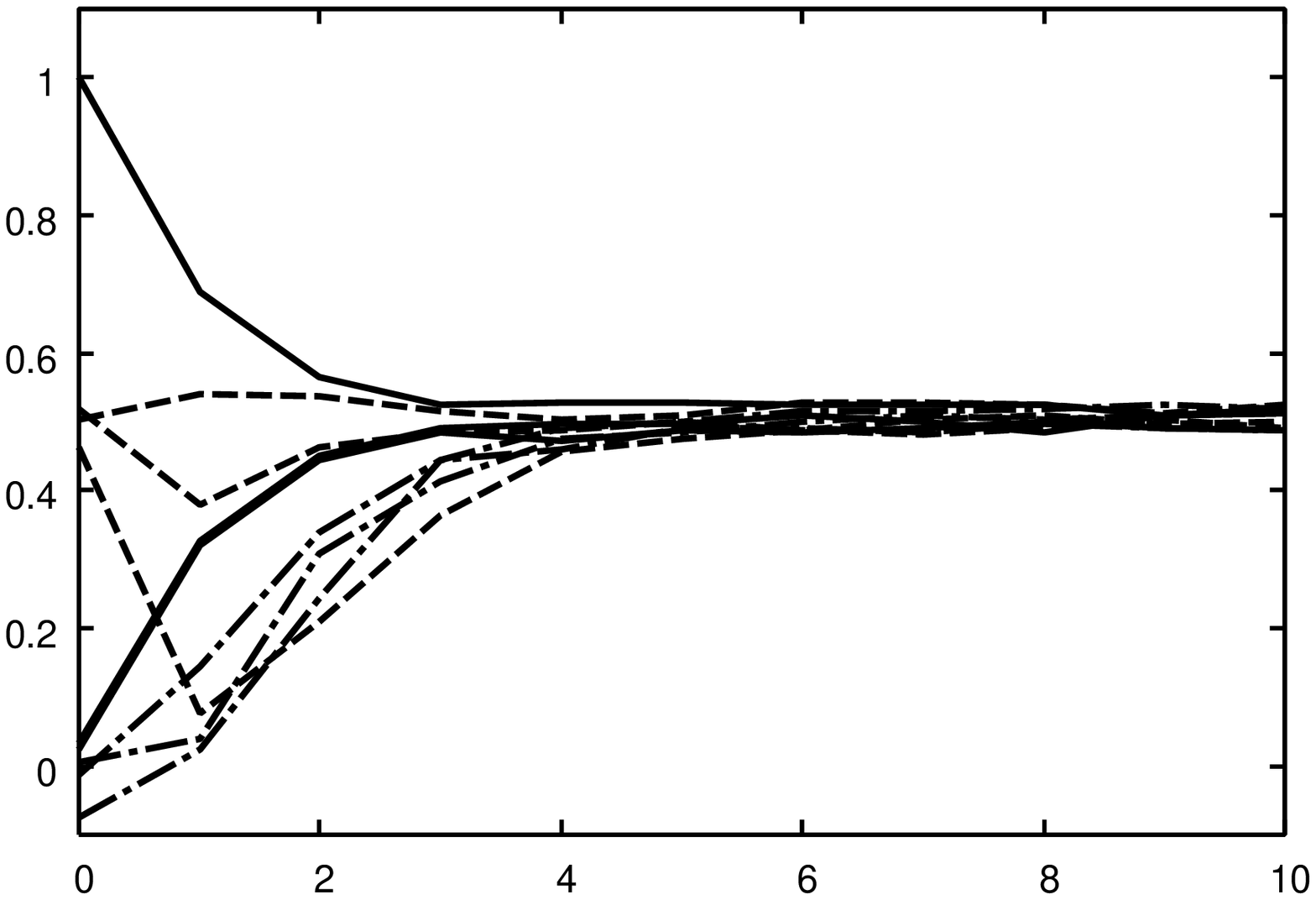}}
\put(112,0){ time[MCS]}
\put(80,40){$m_3$}
\put(80,34){$m_2$}
\put(80,28){$m_1$}
\end{picture}
\caption{Time series of  $m_1 , m_2 , m_3$.
Numerical  results ($N=500$).
$K=1, \mu=1, \tau=1$.
$\tilde{T}=0.1,  \epsilon=1.0, T=0.4$.
Left panel. 
Interaction  $\{J_{ij}^H\}$, in which \H \ appears during
 partial annealing,  is used.
Right panel.  Interaction  $\{J_{ij}^M\}$,  
in which \3M \  appears during
 partial annealing,  is used.
Initial state.
Solid curve: $m_1 =1, m_2=m_3=0$.
Dashed  curve: $m_1 \simeq  m_2  \simeq m_3 \simeq 0.5$.
Dashed dotted curve: random configuration.
} \label{fig.8}
\end{figure}

\setlength{\unitlength}{1.0mm}
\begin{figure}[pthb]
\begin{picture}(130,70)
\put(0,5){\includegraphics[width=8.cm,clip]{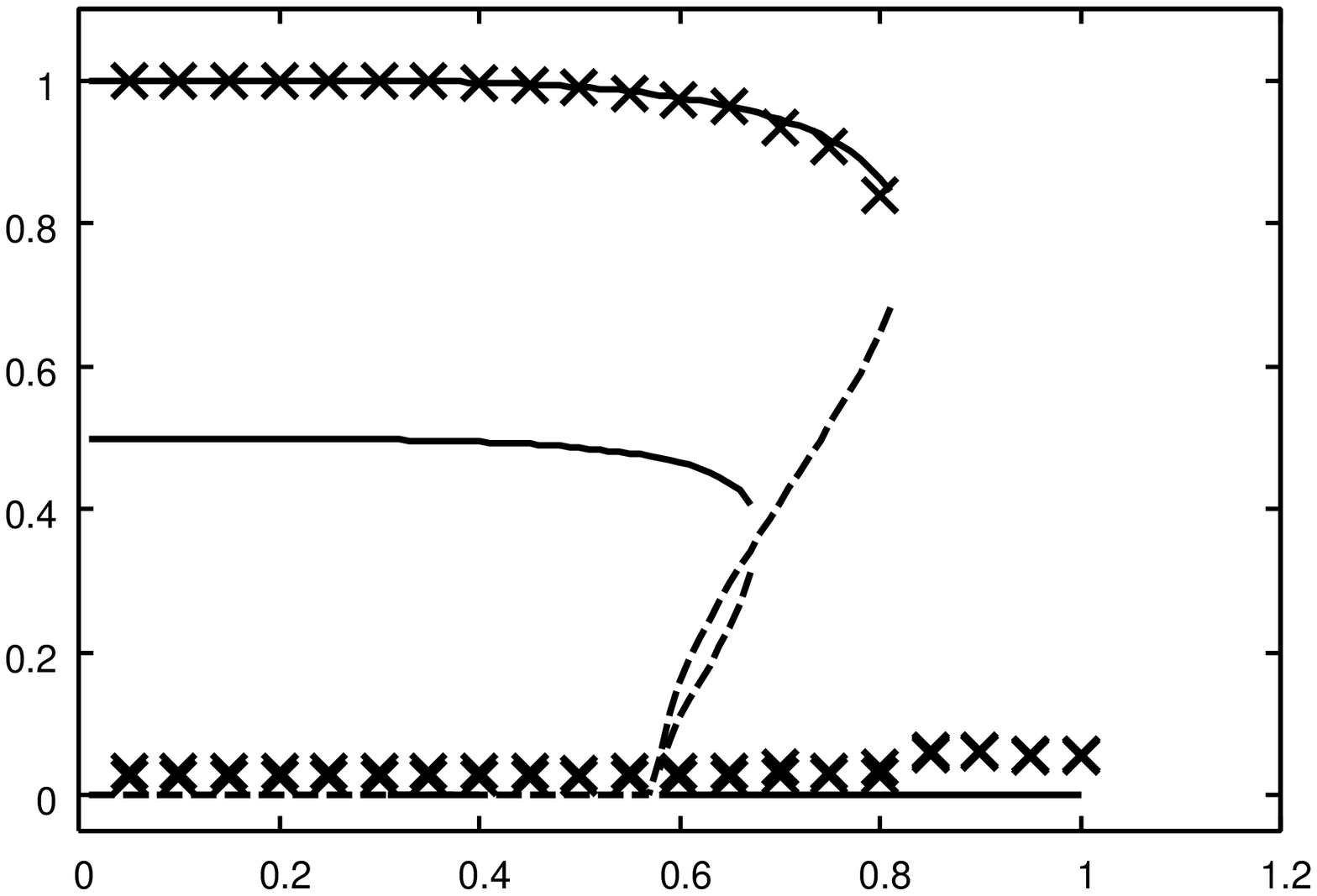}}
\put(43,0){$T$}
\put(-5,40){$m_3$}
\put(-5,34){$m_2$}
\put(-5,28){$m_1$}
\put(85,5){\includegraphics[width=8.cm,clip]{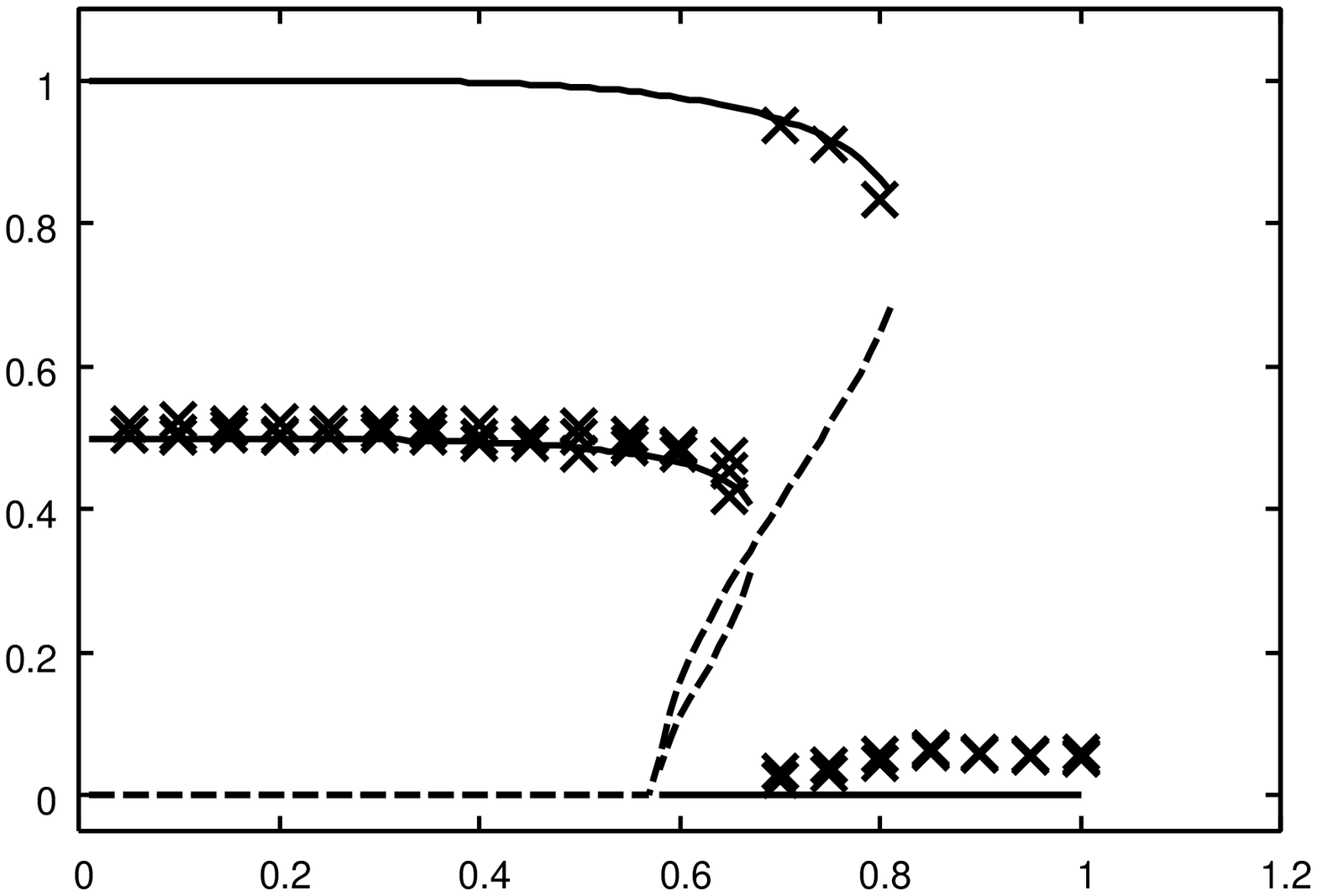}}
\put(120,0){$T$}
\put(80,40){$m_3$}
\put(80,34){$m_2$}
\put(80,28){$m_1$}
\end{picture}
\caption{Temperature dependence of $m_1 , m_2 , m_3$.
$\tilde{T}=0.1,  \epsilon=1.0$.
Curves are theoretical results of   partial
annealing and symbols are simulation results ($N=500$).
Left panel. At each temperature, 
interaction  $\{J_{ij}^H\}$ 
in which \H \ appears during
 partial annealing is used.
Initial state is \3M.
Right panel. At each temperature, 
interaction  $\{J_{ij}^M\}$ 
in which \3M \  appears during
 partial annealing is used.
Initial state is \H.
}\label{fig.9}
\end{figure}
In Fig. \ref{fig.8}, we display the time series of 
$m_{\mu}$ for  $\tilde{T}=0.1, \varepsilon=1.0$ and $T=0.4$.
  We found that when the interaction is 
$\{J_{ij}^H\}$, the neuron system converges to
\H, whereas when 
 the interaction is $\{J_{ij}^M\}$, it  converges to
 \3M, irrespective of initial conditions.
In Fig. \ref{fig.9}, we display the theoretical
and simulation results for the temperature dependence
of $m_1, m_2$ and $ m_3$ for
 $\tilde{T}=0.1$ and $\varepsilon=1.0$.
In the left (right) panel of Fig. \ref{fig.9}, 
at each temperature, the interaction $\{ J_{ij}^H\}$
 ($\{ J_{ij}^M\}$) is used and the initial condition is \3M (\H).
As  seen from these figures,
the resultant attractor by the Monte Carlo simulation
 is \H\ for  $\{J_{ij}^H\}$ and
\3M \ for $\{J_{ij}^M\}$ respectively,
irrespective of the initial conditions.
This implies that  partial annealing has
the effect to enhance the stability of the resultant
attractor and to reduce that of  coexisting attractors.

\section{Summary and discussion}
We investigated the change of the system behaviors 
by  partial annealing
in which the synaptic weights change but much slower than
the neurons. As the basic interaction, we took that of
the Hopfield model, and introduced the coefficient $\varepsilon$
of the Hebbian learning.  
We studied the stationary states of the Langevin equation
by changing parameters, those are 
the learning coefficient  $\varepsilon$, the
neuron ``temperature'' $T$ and the synaptic weight ``temperature''
$\tilde{T}$.\par

First, we studied the phase transition lines  in $(T, \varepsilon)$ 
and $(T, \tilde{T})$ planes and found there is a
tendency that the order of phase transitions of solutions
becomes the second to the first as $\varepsilon$  increases or
$\tilde{T}$ decreases.

Next, by taking into account the AT stability, we drew phase diagrams
in $(T, \varepsilon)$ plane for $\tilde{T}=0.1$ and 0.4.
For $\tilde{T}=0.1$, 
we found that the stable parameter regions for the
Hopfield attractor and the mixed state with 3 patterns
increase as $\varepsilon$ increases.
Further, we found that the ratio of the stable region
of  \3M \  to that of the Hopfield attractor 
increases as $\varepsilon$ increases.
For $\tilde{T}=0.4$, we found that \SG \  appears by the second
order phase transition from \P \  for small $\varepsilon$
and it is the only stable solution.
When $\varepsilon$ is large, the Hopfield attractor  appears by 
the first order phase transition, and the stable region of
the Hopfield attractor increases as $\varepsilon$ increases.  
We confirmed the theoretical results by the numerical
simulations, except for the finite size effects 
observed only for a few set of parameters.\par

In order to study  the nature of interactions generated by partial annealing,
we performed the Monte Carlo simulations using
the synaptic weights obtained by  partial annealing. 
As a result, we found that 
the stability of the attractor which emerges as
the result of  partial annealing is
enhanced and that of the coexistent attractors are reduced.\par
In the present model, the case of $\tilde{T}=0$  and 
$\varepsilon=0$ is nothing but the original Hopfield model.
In this case, the Hopfield attractor and \3M \ are stable for
low temperatures.
We numerically confirmed that the Hopfield attractor 
 is stable for 
all values of $\varepsilon$ when it exists 
for $\tilde{T}=0.1$ although there is a possibility that
it may be AT unstable when $T$ is very low  for $\tilde{T}>0$.
On the other hand, for $\tilde{T}=0.4$
we found the region in which the Hopfield attractor  becomes AT unstable
with the replicon mode eigenvalue $\lambda_3>0$ 
indicating replica symmetry breaking.
This happens when  $\varepsilon$ is less than 1.
Thus, it is concluded that 
if  $\varepsilon$ is larger than some value which depends on 
 $\tilde{T}$, the larger $\varepsilon$ is, 
 the wider the temperature region of the stable Hopfield attractor
  is.\par
From Fig. \ref{fig.5}, we note that for small value of
 $\varepsilon$, partial annealing can not widen
the stability region of the Hopfield attractor.
In particular, it seems that when  $\varepsilon$ becomes negative, 
 i.e. in the case of  unlearning,
no significant change  happens in the phase diagram.
We consider that this is because the loading rate of patterns, $\alpha=
\frac{p}{N}$, is 0 in this study.  If $\alpha$ is positive and large,
 results might change.  This is an interesting unsolved problem.

\appendix

\section{AT stability of solutions}
In this Appendix, the eigenvalues and eigenvectors
of the Hessian are calculated.
In the below, the  symbols $\alpha, \beta, \gamma$
and $\delta$ indicate the replica indexes and
symbols  $\mu$ and $\nu$ do the pattern induces.\\
Let the Hessian matrix  ${\cal G}$ be the $L \times L$ matrix.
$L$ is the summation of the dimensions of
$\{ \dsp \epsilon^\alpha_\nu \}$ space
and that of $\{\dsp \eta^{\alpha\beta}\}$ space, i.e., 
$\dsp L=  3n+{}_n C_2 =\frac{n(n+5)}{2}$ for $p=3$.
The eigenvalue equation is expressed as
\begin{eqnarray}
{\cal G}\bfmu=\lambda\bfmu, \label{eq:koyuuchieq}
\end{eqnarray}
where $\lambda$ is an eigenvalue of 
 ${\cal G}$ and $\bfmu$ is the eigenvector belonging to  $\lambda$.
 $\bfmu$ takes the following form:
\begin{center}
\begin{equation}
\bfmu = 
\left(
\begin{array}{c}
\{\epsilon_\nu ^\alpha\} \\
\{\eta^{\alpha\beta}\}
\end{array}
\right). \label{eq:koyuuvector}
\end{equation}
\end{center}
$\dsp \{\epsilon^\alpha_\nu \}$ and $\dsp \{\eta^{\alpha\beta}\}$
denote a $3n$ dimensional 
and a ${}_n C_2$ dimensional 
column vectors, respectively.
In the below, $[\cdots]$ is the average over
patterns $\xi$ and  a bar denotes the following average:
\begin{eqnarray}
\overline{f(\Xi)} &\equiv& \Omega^{-1}\int Dx\cosh^n(\Xi)f(\Xi),
 \label{eq:fbar}\\
\Omega &\equiv& \int Dx\cosh^n(\Xi), \label{eq:Omega}\\
\Xi &=&
 \beta\bigg\{\sqrt{\frac{q}{\mu\tilde{\beta}}}x
+\frac{K}{\mu\sqrt{p}}\sum_\nu m_\nu\xi^\nu\bigg\}
= \sqrt{\kappa q}x
+\beta J \sum_\nu m_\nu\xi^\nu,\\
&& \kappa \equiv \frac{\beta ^2}{\mu \tilde{\beta}}, 
\ J \equiv \frac{K}{\mu\sqrt{p}}.
\end{eqnarray}
Further, $\bra\cdots\ket$ denotes the following
average at the replica symmetric solution. 
\begin{eqnarray*}
\bra\cdots\ket &=& \frac{{\rm Tr}_{\bfsigma}e^{\tilde{H}}\cdots}
{{\rm Tr}_{\bfsigma} e^{\tilde{H}}},\\
\tilde{H} &=&
 \kappa\sum_{\alpha<\beta}(q_{\alpha\beta})^2\sigma^\alpha\sigma^\beta
+\beta J\sum_{\alpha\nu}m_\nu^\alpha\sigma^\alpha\xi^\nu \\
 &=& \kappa q^2\sum_{\alpha<\beta}\sigma^\alpha\sigma^\beta
+\beta J\sum_{\alpha\nu}m_\nu\sigma^\alpha\xi^\nu.
\end{eqnarray*}
\subsection{Hopfield attractor}
We put $m_1=m, m_2=m_3=0$. Then, the non-zero
elements of ${\cal G}$ are the 7 quantities defined as
\begin{eqnarray}
{\cal G}_{({\alpha}{\nu})({\alpha}{\nu})}&\equiv&
\frac{\partial^2 G}{\partial m_\nu ^{\alpha 2}}=
-\beta J(1-{\beta}J(1-m^2))=A,\label{eq:HAG1}\\
{\cal G}_{({\alpha}{\nu})({\beta}{\nu})} &\equiv&
\frac{\partial ^2 G}{\partial m^\alpha_\nu\partial m^ \beta_\nu} =
(\beta J)^2 (q-m^2) = B,  \qquad (\alpha\ne\beta), \label{eq:HAG2}\\
{\cal G}_{({\alpha}{\beta})({\alpha}{\beta})} &\equiv&
\frac{\partial ^2 G}{\partial q_{\alpha\beta}\partial q_ {\alpha\beta}} 
= -\kappa[1-\kappa(1-\bra \sigma^\alpha\sigma^\beta \ket ^2)] \nonumber\\
& &  \hspace{1.7cm} = -\kappa[1-\kappa(1-q^2)]
= P,  \qquad  (\alpha\ne\beta). \label{eq:HAG3}\\
{\cal G}_{({\alpha}{\beta})({\alpha}{\gamma})} &\equiv&
\frac{\partial ^2 G}{\partial q_{\alpha\beta}\partial q_ {\alpha\gamma}}
=\kappa^2[\bra\sigma^\beta\sigma^\gamma\ket-\bra\sigma^\alpha\sigma^\beta\ket^2
] \nonumber\\
& & \hspace{1.7cm} =\kappa^2(q-q^2)
= Q, \mbox{
$\alpha, \beta$ and $\gamma$ are all different.
}
\label{eq:HAG4}\\
{\cal G}_{({\alpha}{\beta})({\gamma}{\delta})} &\equiv&
\frac{\partial ^2 G}{\partial q_{\alpha\beta}\partial q_ {\gamma\delta}}
=\kappa^2[\bra\sigma^\alpha\sigma^\beta\sigma^\gamma\sigma^\delta\ket
-\bra\sigma^\alpha\sigma^\beta\ket^2]
 =\kappa^2(\overline{\tanh^4\Xi_1}-q^2)
= R, \label{eq:HAG5}\nonumber\\
& & \hspace{3.5cm}\mbox{ 
$\alpha, \beta, \gamma$ and $\delta$ are all different.}
\end{eqnarray}
\begin{eqnarray}
{\cal G}_{({\alpha}{\beta})({\alpha}{\nu})} &\equiv&
\frac{\partial ^2 G}{\partial q_{\alpha\beta}\partial m_\nu^\alpha}=
\kappa\beta J m (1-q)\delta_{\nu 1} = C\delta_{\nu 1},  (\alpha\ne\beta).\label{eq:HAG6}\\
{\cal G}_{({\beta}{\gamma})({\alpha}{\nu})} &\equiv&
\frac{\partial ^2 G}{\partial q_{\beta\gamma}\partial m_\nu^\alpha}=
\kappa\beta J 
(\overline{\tanh^3\Xi_1}-qm)\delta_{\nu 1} = D\delta_{\nu 1}, \nonumber\\
& & \hspace{3.5cm} (\alpha\ne\beta , \alpha\ne\gamma)\label{eq:HAG7}.
\end{eqnarray}
Here, $\dsp \Xi_1=\sqrt{\kappa q}x+\beta Jm$.
We list the eigenvalues, their degeneracies and eigenvectors.
\begin{eqnarray}
\lambda_1^{(1)\pm} &=& \frac{1}{2}\{X\pm\sqrt{Y^2+Z}\},
\ \mbox{ degeneracy: 1 for each,}\\
&&X =  A+(n-1)B+P+2(n-2)Q+\frac{(n-2)(n-3)}{2}R, \nonumber \\
&&Y =  A+(n-1)B-P-2(n-2)Q-\frac{(n-2)(n-3)}{2}R, \nonumber \\
&&Z =  2(n-1)\{2C+(n-2)D\}^2,\nonumber \\
&& \epsilon^\alpha_1=a, \epsilon^\alpha_2=0,
\epsilon^\alpha_3=0, \eta^{\alpha\beta}=b,\nonumber \\
\lambda_1^{(2)}&=&A+(n-1)B,
\ \mbox{ degeneracy: 2},\\
&& \epsilon^{\alpha}_1=0, \epsilon^{\alpha}_{\mu}=a^\prime,
  \eta^{\alpha \beta}=0, \hspace{1cm} \mbox{$\mu =2$ or 3 },\nonumber \\
\lambda_2^{(1)\pm} &=& 
\frac{1}{2} \{ X^\prime \pm \sqrt{(Y^\prime) ^2+Z^\prime}\}, 
\ \mbox{ degeneracy: $(n-1)$ for each,}\\
&& X^\prime = A-B+P+(n-4)Q-(n-3)R,\nonumber \\
&& Y^\prime = A-B-P-(n-4)Q+(n-3)R, \nonumber \\
&& Z^\prime = 4(n-2)(C-D)^2, \nonumber \\
&&  \epsilon^\theta_1=c_1, \ \epsilon^\alpha_1=d_1,\
\epsilon^\beta_2=0, \ \epsilon^\beta_3=0, 
\ \mbox{$\alpha \ne \theta$, $\theta$ is some replica index}, \nonumber \\
&& \eta^{\theta \beta}=\eta^{\alpha \theta}=f,
\eta^{\alpha \beta}=g, \hspace{1cm}
 \mbox{$\alpha \ne \theta$ and $\beta \ne \theta$},\nonumber \\
\lambda_2^{(2)}& = & A-B,
\ \mbox{ degeneracy: $2(n-1)$},\\
&&  \epsilon^{\theta}_\mu=c_2, \ \epsilon^{\alpha}_\mu=d_2,\
\epsilon^\beta_{\nu}=0, 
\ \mbox{$\alpha \ne \theta$, $\nu \ne \mu$,  $\theta$ is some replica index
, $\mu =2$ or 3}, \nonumber \\
&& \eta^{\alpha \beta}=0, \ \mbox{ for any $\alpha, \beta$}, \nonumber \\
\lambda_3 & = & P-2Q+R, \ \mbox{ degeneracy: $ \frac{n(n-1)}{2}-n$},\\
&& \epsilon^{\theta_1}_1= \epsilon^{\theta_2}_1=h_1, 
\epsilon^{\alpha}_1=h_2, \epsilon^{\beta}_{\nu}=0,
\theta_1 \ne \theta_2, \alpha \ne \theta_1,
\alpha \ne \theta_2,  \nu \ne 1,\nonumber\\
&& \mbox{  $\theta_1$ and $\theta_2$ are some replica indexes},\nonumber\\
&& \eta^{\theta_1 \theta_2}=u, 
\eta^{\theta_1 \alpha}=\eta^{\theta_2 \alpha }=v,
\eta^{\alpha \beta}=w, \nonumber\\
&& \hspace{3.5cm}\mbox{$\alpha \ne \theta_1, \alpha \ne \theta_2$ and
$\beta \ne \theta_1, \beta \ne \theta_2$}.
\end{eqnarray}
\subsection{Mixed state with 3 patterns}
We put $m_1=m_2=m_3=m$.
$\Xi_m$ is defined by
\begin{eqnarray}
\Xi &=& \sqrt{\kappa q}x+\beta J\sum_\nu m_\nu\xi_\nu
= \sqrt{\kappa q}x+\beta Jm(\xi_1+\xi_2+\xi_3)\equiv \Xi_m.
\end{eqnarray}
Now, we list the non-zero elements of ${\cal G}$.
\begin{eqnarray}
{\cal G}_{({\alpha}{\nu})({\alpha}{\nu})} \equiv
\frac{\partial ^2 G}{\partial m_\nu ^{\alpha 2}} =
-\beta J+(\beta J)^2\Big\{1-\big[(\overline{\tanh\Xi_m})^2\big]\Big\} = A_1.
\end{eqnarray}
Since this quantity does not depend on $\nu$, we define it $A_1$.
In the below, we assume $\mu \ne \nu$ and 
$\alpha, \beta, \gamma$ and $\delta$ are all different.
\begin{eqnarray}
{\cal G}_{({\alpha}{\nu})({\alpha}{\mu})} &\equiv&
\frac{\partial ^2 G}{\partial m^{\alpha}_{\nu} \partial
 m^{\alpha}_{\mu}} 
= (\beta
J)^2\Big\{\big[\bra\sigma^\alpha\sigma^\beta\xi^\nu\xi^\mu\ket\big]
-\big[\bra\sigma^\alpha\xi^\nu\ket
 \bra\sigma^\alpha\xi^\mu\ket\big]\Big\} \nonumber \\
 &=& (\beta  J)^2\Big\{\big[\xi^\nu\xi^\mu\big]
-\big[\bra\sigma^\alpha\ket^2\xi^\mu\xi^\nu\big]\Big\} \nonumber \\
 &=& -(\beta  J)^2\Big[(\overline{\tanh\Xi_m})^2\xi^\mu\xi^\nu\Big] 
 = A_2,\\
{\cal G}_{({\alpha}{\nu})({\beta}{\nu})} &\equiv&
\frac{\partial ^2 G}{\partial m^\alpha_\nu\partial m^ \beta_\nu} 
= (\beta J)^2\Big\{\big[\bra\sigma^\alpha\sigma^\beta\ket\big]
-\big[\bra\sigma^\alpha\ket\xi^\nu\bra\sigma^\beta
\ket\xi^\nu\big]\Big\}
  \nonumber \\
& = & (\beta J)^2\Big\{\big[\overline{\tanh^2\Xi_m}\big]
-\big[(\overline{\tanh\Xi_m})^2\big]\Big\}  \nonumber \\
&= & (\beta J)^2\Big\{q-\big[(\overline{\tanh\Xi_m})^2\big]\Big\}
= B_1.
\end{eqnarray}
At the last line, we used the following relations,
\begin{eqnarray}
q&=&\Big[\bra\sigma^\alpha\sigma^\beta\ket\Big]
=\Big[\;\overline{\tanh^2\Xi_m}\;\Big],\\
m_\nu^\alpha&=&\Big[\bra\sigma^\alpha\ket\xi^\nu\Big]
=\Big[\overline{\tanh\Xi_m}\xi^\nu\Big]=m_\nu=m.
\end{eqnarray}
\begin{eqnarray}
{\cal G}_{({\alpha}{\nu})({\beta}{\mu})} &\equiv&
\frac{\partial ^2 G}{\partial m^\alpha_\nu\partial m^ \beta_\mu} 
= (\beta
 J)^2\Big\{\big[\bra\sigma^\alpha\sigma^\beta\ket\xi^\nu\xi^\mu\big]
-\big[\bra\sigma^\alpha\ket\bra\sigma^\beta\ket\xi^\nu\xi^\mu\big]\Big\} 
\nonumber \\
&=& (\beta J)^2\Big\{\overline{\tanh^2\Xi_m}\xi^\nu\xi^\mu
-\big[(\overline{\tanh\Xi_m})^2\xi^\mu\xi^\nu\big]\Big\} = B_2,\\
{\cal G}_{(\alpha\beta)(\alpha\nu)} &\equiv&
 \frac{\partial^2 G}{\partial q_{\alpha\beta}\partial m_\nu^\alpha} 
= \kappa\beta
J\Big[\bra\sigma^\alpha\sigma^\beta\sigma^\alpha\xi^\nu\ket
-\bra\sigma^\alpha\sigma^\beta\ket\bra\sigma^\alpha\xi^\nu\ket\Big]\nonumber \\
&= & \kappa\beta
 J\bigg\{\Big[\overline{\tanh\Xi_m}\xi^\nu\Big]
-\big[\overline{\tanh^2\Xi_m}\cdot
\overline{\tanh\Xi_m}\xi^\nu\big]\bigg\}\nonumber \\
& = & \kappa\beta
 J\Big\{m-\big[\overline{\tanh^2\Xi_m}\cdot
\overline{\tanh\Xi_m}\xi^\nu\big]\Big\}
\equiv C,\\
{\cal G}_{(\alpha\beta)(\nu\gamma)} &\equiv& 
\frac{\partial^2 G}{\partial q_{\alpha\beta}\partial m_\nu^\gamma}
= \kappa\beta
 J\Big[\bra\sigma^\alpha\sigma^\beta\sigma^\gamma\ket\xi^\nu
-\bra\sigma^\alpha\sigma^\beta\ket\bra\sigma^\gamma\ket\xi^\nu\Big]\nonumber\\
&=& \kappa\beta
 J\Big\{\big[\overline{\tanh^3\Xi_m}\xi^\nu\big]
-\big[\overline{\tanh^2\Xi_m}\cdot\overline{\tanh\Xi_m}\xi^\nu\big]\Big\}
\equiv D,\\
{\cal G}_{({\alpha}{\beta})({\alpha}{\beta})} &\equiv&
\frac{\partial ^2 G}{\partial q_{\alpha\beta}^2} 
=
-\kappa+\kappa^2\Big\{1-\big[\bra\sigma^\alpha\sigma^\beta\ket^2\big]\Big\}
\nonumber \\
& = & -\kappa+\kappa^2\Big\{1-\big[(\overline{\tanh^2\Xi_m})^2\big]\Big\}
\equiv P,\\
{\cal G}_{({\alpha}{\beta})({\alpha}{\gamma})} &\equiv&
\frac{\partial ^2 G}{\partial q_{\alpha\beta}\partial q_ {\alpha\gamma}}
=\kappa^2\big[\bra\sigma^\alpha\sigma^\beta\sigma^\alpha\sigma^\gamma\ket
-\bra\sigma^\alpha\sigma^\beta\ket^2\big]
\nonumber \\
&=& \kappa^2\Big\{\big[\overline{\tanh^2\Xi_m}\big]
-\big[(\overline{\tanh^2\Xi_m})^2\big]\Big\}
\nonumber \\
& =& \kappa^2\Big\{q-\big[(\overline{\tanh^2\Xi_m})^2\big]\Big\} 
\equiv Q, \\
{\cal G}_{({\alpha}{\beta})({\gamma}{\delta})} &\equiv&
\frac{\partial ^2 G}{\partial q_{\alpha\beta}\partial q_ {\gamma\delta}}
=\kappa^2\big[\bra\sigma^\alpha\sigma^\beta\sigma^\gamma\sigma^\delta\ket
-\bra\sigma^\alpha\sigma^\beta\ket^2\big]
\nonumber \\
& = & \kappa^2\Big\{\big[\overline{\tanh^4\Xi_m}\big]
-\big[(\overline{\tanh^2\Xi_m})^2\big]\Big\}\equiv R.
\end{eqnarray}
We list the eigenvalues, their degeneracies and eigenvectors.
\begin{eqnarray}
\lambda_1^{(1)\pm} &=& \frac{1}{2}\{X\pm\sqrt{Y^2+Z}\},
\ \mbox{ degeneracy: 1 for each,}\\
&&X =  A+(n-1)B+P+2(n-2)Q+\frac{(n-2)(n-3)}{2}R, \nonumber \\
&&Y =  A+(n-1)B-P-2(n-2)Q-\frac{(n-2)(n-3)}{2}R, \nonumber \\
&&Z =  6(n-1)\big\{2C+(n-2)D\big\}^2,\nonumber \\
&&A =   A_1+(n-1)B_1, \ B=\frac{2(A_2+(n-1)B_2)}{n-1}, \nonumber \\
&& \epsilon^\alpha_1= \epsilon^\alpha_2=\epsilon^\alpha_3=a,
 \eta^{\alpha\beta}=b,\nonumber \\
\lambda_1^{(2)} & = &A_1-A_2+(n-1)(B_1-B_2), \ \mbox{ degeneracy: 2},\\
&& \epsilon^{\alpha}_1=a_1^{\prime}, \epsilon^{\alpha}_2=a_2^\prime,
\epsilon^{\alpha}_3=a_3^\prime, \ a_1^\prime+a_2^\prime+ a_3^\prime=0,\
  \eta^{\alpha \beta}=0,\nonumber \\
\lambda_2^{(1)\pm} &=&
 \frac{1}{2}\{X^\prime\pm\sqrt{(Y^\prime)^2+Z^\prime}\}, 
\ \mbox{ degeneracy: $(n-1)$ for each,}\\
&& X^\prime = A_1-B_1+2(A_2-B_2)+P+(n-4)Q-(n-3)R, \nonumber \\
&& Y^\prime = A_1-B_1+2(A_2-B_2)-P-(n-4)Q+(n-3)R, \nonumber \\
&& Z^\prime = 12(n-2)(C-D)^2,\nonumber \\
&& \epsilon^\theta_1=c_1, \ \epsilon^\alpha_1=d_1, 
\epsilon^\theta_2=c_2, \ \epsilon^\alpha_2=d_2, 
\epsilon^\theta_3=c_3, \ \epsilon^\alpha_3=d_3, \nonumber \\
&& \hspace{3.5cm}
 \mbox{$\alpha \ne \theta$, $\theta$ is some replica index},\nonumber \\
&& \eta^{\theta \beta}=\eta^{\alpha \theta}=, 
\eta^{\alpha \beta}=g, \hspace{1cm}
 \mbox{$\alpha \ne \theta$ and $\beta \ne \theta$},\nonumber \\
\lambda_2^{(2)}&=&A_1-A_2-(B_1-B_2),
\ \mbox{ degeneracy: $2(n-1)$}, \\
&& \epsilon^\theta_1=c_1, \ \epsilon^\alpha_1=d_1, 
\epsilon^\theta_2=c_2, \ \epsilon^\alpha_2=d_2, 
\epsilon^\theta_3=c_3, \ \epsilon^\alpha_3=d_3, \nonumber \\
&& \hspace{3.5cm}
\ \mbox{$\alpha \ne \theta$, $\theta$ is some replica index},\nonumber \\
&& \eta^{\alpha \beta}=0,\hspace{1cm}
\ \mbox{ for any $\alpha, \beta$},\nonumber \\
&& \eta^{\alpha \beta}=0, \ \mbox{ for any $\alpha, \beta$}, \nonumber
 \\
&& c_i=(1-n)d_i, \ d_1+d_2+ d_3=0,\nonumber \\
\lambda_3 & = & P-2Q+R, \ \mbox{ degeneracy: $ \frac{n(n-1)}{2}-n$},\\
&& \epsilon^{\theta_1}_{\mu}= \epsilon^{\theta_2}_{\mu}=r_{\mu},
\ \epsilon^{\alpha}_{\mu}=s_{\mu},\ \ \mu=1,2,3,
\theta_1 \ne \theta_2, \ \alpha \ne \theta_1,
\alpha \ne \theta_2, \nonumber\\
&& \mbox{  $\theta_1$ and $\theta_2$ are some replica indexes},\nonumber\\
&& \eta^{\theta_1 \theta_2}=u, \
\eta^{\theta_1 \alpha}=\eta^{\theta_2 \alpha }=v,\
\eta^{\alpha \beta}=w, \nonumber \\
&& \hspace{3.5cm}
 \mbox{$\alpha \ne \theta_1, \alpha \ne \theta_2$ and
$\beta \ne \theta_1, \beta \ne \theta_2$}. \nonumber
\end{eqnarray}

\subsection{Spin Glass}
For the spin glass solution,
$m_{\mu}=0$ and  $q \ne 0$.
Thus, we only have to put $m=0, C=D=0 $
in the quantities for the Hopfield attractor.
Thus, we obtain the following eigenvalues:
\begin{eqnarray}
\lambda_1^{(1) +} &=& A+(n-1)B=
\lambda_1^{(2)} =\lambda_1^{(3)}, \\
\lambda_1^{(1) -} &=& P+2(n-2)Q+\frac{(n-2)(n-3)}{2}R, \\
\lambda_2^{(1)+} &=& A-B,\\
\lambda_2^{(1)-} &=&  P+(n-1)Q-(n-3)R=
\lambda_2^{(2)} =\lambda_2^{(3)}, \\
\lambda_3 &=&  P-2Q+R.
\end{eqnarray}

\end{document}